\documentclass[12pt]{article}
\usepackage[utf8]{inputenc}
\usepackage[english]{babel}
\usepackage[top = 2 cm, bottom = 2 cm, left = 2.5 cm, right = 1.5 cm]{geometry}
\usepackage[]{amsmath}
\usepackage[pdftex]{graphicx}
\graphicspath{ {images/} }
\usepackage{float}
\restylefloat{figure} 
\usepackage{subfig}
\usepackage{indentfirst}
\usepackage{amsthm}
\usepackage{amsfonts}
\usepackage{mathrsfs}
\usepackage{mathtools}
\usepackage{cmap}
\usepackage{exercise}
\usepackage{xcolor}
\usepackage{breqn}
\usepackage{cancel}
\usepackage{amsfonts}
\usepackage{comment}
\usepackage{tcolorbox}
\usepackage[pdftex,unicode,
colorlinks=true,
linkcolor = blue]{hyperref}
\usepackage[offset=1.2em]{simpler-wick}
\usepackage[compat=1.1.0]{tikz-feynman}
\usepackage{tikz-3dplot}
\usetikzlibrary{arrows}
\usetikzlibrary{arrows.spaced}
\usetikzlibrary{decorations.pathmorphing}
\usetikzlibrary{snakes}
\usetikzlibrary{shadows}
\usetikzlibrary{decorations.pathreplacing,decorations.markings}
\usetikzlibrary {fadings,patterns}
\usepackage[bulletsep]{collref}

\def\nullify#1{}

\makeatletter
\let\old@makecaption=\@makecaption
\def\@makecaption{\small\old@makecaption}
\makeatother

\usepackage{wasysym}
\usepackage{amssymb}

\usepackage[colorinlistoftodos]{todonotes}%allows to do colorful comments
\newcommand{\mathbbm}[1]{\text{\usefont{U}{bbm}{m}{n}#1}}

\usepackage{marginnote}

\definecolor{nicklethistwink}{rgb}{0.69, 0.19, 0.38}
\definecolor{ticklemepink}{rgb}{0.99, 0.54, 0.67}
\definecolor{mistyrose}{rgb}{1.0, 0.89, 0.88}
\definecolor{lolololol}{rgb}{0.25, 0.5, 0.75}
\definecolor{upsdellred}{rgb}{0.68, 0.09, 0.13}
\definecolor{champagne}{rgb}{0.97, 0.91, 0.81}
\definecolor{ghostwhite}{rgb}{0.97, 0.97, 1.0}
\definecolor{ivory}{rgb}{1.0, 1.0, 0.94}
\definecolor{bubblegum}{rgb}{0.99, 0.76, 0.8}
\definecolor{mintcream}{rgb}{0.96, 1.0, 0.98}
\definecolor{honeydew}{rgb}{0.94, 1.0, 0.94}
\definecolor{magnolia}{rgb}{0.97, 0.96, 1.0}
\definecolor{isabelline}{rgb}{0.96, 0.94, 0.93}
\definecolor{bubbles}{rgb}{0.91, 1.0, 1.0}
\definecolor{floralwhite}{rgb}{1.0, 0.98, 0.94}
\definecolor{turquoise}{rgb}{0.19, 0.84, 0.78}
\definecolor{bittersweet}{rgb}{1.0, 0.44, 0.37}
\definecolor{ultramarineblue}{rgb}{0.25, 0.4, 0.96}
\def\fn#1{\footnote{#1}}

\def\eqref#1{(\ref{#1})}
\def\comma{\,,}
\def\period{\,.}

\def\ket#1{|#1\rangle}

\numberwithin{equation}{section}

\def\beq{\begin{equation}}
\def\eeq{\end{equation}}

\newif{\ifremarks}

\ifremarks\newcommand{\remarkkz}[1]{%
  {\renewcommand{\bfdefault}{b}\color[RGB]{0,150,0}{\textbf{K:~#1}}}}\fi
\providecommand{\remarkkz}[1]{\ignorespaces}

\ifremarks\newcommand{\remarksk}[1]{%
  {\renewcommand{\bfdefault}{b}\color[RGB]{0,0,150}{\textbf{S:~#1}}}}\fi
\providecommand{\remarksk}[1]{\ignorespaces}

\ifremarks\newcommand{\remarkfc}[1]{%
  {\renewcommand{\bfdefault}{b}\color[RGB]{150,0,150}{\textbf{F:~#1}}}}\fi
\providecommand{\remarkfc}[1]{\ignorespaces}

\begin{document}
	\title{\vspace{0.1cm}{\Large {\bf 
				 Coulomb Branch and Integrability}}}\vspace{10pt}
		\author{Frank Coronado$^{1}$,\,\, Shota Komatsu$^{2,3}$,\,\, Konstantin Zarembo$^{4,5}$\vspace{10pt}\\		
			{}\date{ }
		}
	\maketitle
	
	\vspace{-5.9cm}
	
	\begin{center}
		\hfill \\
	\end{center}
	
	\vspace{3.0cm}
	
	\begin{center}
		
		 {\small {\it $^1$ Institut f\"{u}r Theoretische Physik, ETH Z\"{u}rich, CH-8093 Z\"{u}rich, Switzerland}}\\
 {\small {\it $^2$ Department of Theoretical Physics, CERN, 1211 Meyrin, Switzerland}}\\ 
 {\small {\it $^3$ Blackett Laboratory, Imperial College London, SW7 2AZ London, United Kingdom}}\\ 
 {\small {\it $^4$Nordita, KTH Royal Institute of Technology and Stockholm University,}}\\
 {\small {\it Hannes Alfv\'ens v\"ag 12, SE-106 91 Stockholm, Sweden}}\\
   {\small {\it $^5$ Niels Bohr Institute, Copenhagen University, Blegdamsvej 17, 2100 Copenhagen, Denmark}}

	\end{center}
	
	\vspace{1cm}
	
	\begin{abstract}
		We study one-point functions of non-BPS single-trace operators on the Coulomb branch of planar $\mathcal{N}=4$ supersymmetric Yang-Mills theory. Holography relates them to overlaps between on-shell closed string states and a boundary state describing a probe D3-brane in $AdS_5\times S^{5}$. Assuming that the D-brane preserves integrability, we formulate and solve integrable bootstrap equations satisfied by the boundary state at finite ’t Hooft coupling. This leads to a closed-form determinant expression for one-point functions at finite coupling, valid for sufficiently long operators. We test the result against direct field theory computations at tree level and one loop, finding perfect agreement.
	\end{abstract}
	
	\vspace{.5cm}  
	
	\newpage
	
	\tableofcontents
	\setlength{\parskip}{0.5em}
	\section{Introduction}
    Understanding the strongly-coupled dynamics of gauge theories remains a major challenge in theoretical physics. Over the last decades, notable progress has been made in planar $\mathcal{N}=4$ supersymmetric Yang-Mills (SYM) theory in four dimensions, a highly symmetric theory that serves as a testing ground for non-perturbative techniques. This progress is largely due to its connection with integrability \cite{Beisert:2010jr} and holography \cite{Maldacena:1997re}: In \cite{Minahan:2002ve}, it was shown that computing the conformal dimensions of single-trace operators maps to determining the energy spectrum of certain spin chains which are integrable and solvable via the Bethe ansatz. In addition, $\mathcal{N}=4$ SYM is conjectured to be dual to type IIB superstring theory in $AdS_5\times S^{5}$, and the corresponding string sigma model in this background is classically integrable \cite{Bena:2003wd}. Integrability, which appears on both sides of the duality, is now believed to extend beyond the weak-coupling or classical regimes, allowing for exact non-perturbative computations at  finite 't Hooft coupling. 
    
    Despite these advances, $\mathcal{N}=4$ SYM differs in key ways from realistic gauge theories. In particular, it is a {\it conformal} theory and does not generate mass scales via dimensional transmutations as in QCD. This limits its direct phenomenological relevance.
    One way to bridge this gap, while preserving the tractability due to  integrability and holography, is to consider the theory on its moduli space of vacua. In this paper, we continue the analyses in \cite{Ivanovskiy:2024vel} and study $\mathcal{N}=4$ SYM on the Coulomb branch. More specifically, we consider the vacuum configuration in which the gauge group is broken as $U(N)\to U(1) \times U(N-1)$ by a Higgs condensate
    	\begin{equation}\label{higgsvev}
		\phi^{\rm cl}_i =
		\left(\begin{matrix}
			v_i & 0 & \ldots & 0 \\
			0 & 0 &  &0 \\
			\vdots &  & \ddots &0 \\
			0 & 0 & 0 &0
		\end{matrix}\right) \in U(N), \qquad  \sum\limits_{i=1}^6 v_i^2=v^2 \comma
	\end{equation}
    in the planar limit of the large-$N$ expansion with the 't~Hooft coupling  parameterized as
\begin{equation}
    \lambda=g_{\rm YM}^2N, \qquad g=\frac{\sqrt{\lambda}}{4\pi}\period
\end{equation}
    The Higgs condensate \eqref{higgsvev} introduces the mass scale and breaks the conformal symmetry, bringing the theory closer to realistic theories. It also introduces richer dynamics, such as the vacuum condensates (i.e.~one-point functions) of local operators \cite{Ivanovskiy:2024vel,Cuomo:2024vfk,Cuomo:2024fuy} and massive bound states of W-bosons \cite{Caron-Huot:2014gia}. Furthermore, several lines of indirect evidence suggest that integrability persists on the Coulomb branch given by \eqref{higgsvev}. These include results from string theory \cite{Dekel:2011ja}, perturbative analyses of scattering amplitudes \cite{Alday:2009zm,Loebbert:2020hxk,Loebbert:2020tje,Arkani-Hamed:2023epq} and correlation functions \cite{Caron-Huot:2021usw}, and studies of the W-boson bound-state spectrum \cite{Caron-Huot:2014gia}. 
    
    In this work, we take a direct step forward by deriving a closed-form expression for one-point functions of single-trace operators at finite ’t Hooft coupling, leveraging integrability. This expression, known as the {\it asymptotic formula}, is expected to be valid for sufficiently long operators, up to corrections from so-called wrapping effects.

Our approach builds on successful strategies developed in related contexts, such as defect conformal field theories \cite{deLeeuw:2015hxa,Buhl-Mortensen:2015gfd,Buhl-Mortensen:2017ind,Komatsu:2020sup,Gombor:2020kgu,Gombor:2020auk,Kristjansen:2023ysz,Kristjansen:2024map,Holguin:2025bfe,Chalabi:2025nbg} and heavy-heavy-light three-point functions \cite{Jiang:2019xdz,Jiang:2019zig}. In particular, we follow the integrable bootstrap program laid out in \cite{Ghoshal:1993tm,Komatsu:2020sup,Jiang:2019xdz,Gombor:2020kgu,Gombor:2020auk}. The central insight is that, in the holographic dual description, the vacuum condensates correspond to overlaps between on-shell closed string states and a boundary state representing a probe D3-brane.
Assuming that this boundary state preserves integrability, we formulate a set of consistency conditions\footnote{These boundary integrable bootstrap equations were initially formulated by Ghoshal and Zamolodchikov for general relativistic integrable QFT in \cite{Ghoshal:1993tm}.} that constrain the boundary reflection matrix. We then provide a solution to these conditions at finite ’t Hooft coupling and derive the asymptotic formula for one-point functions in closed form, which has the following structure:
\begin{equation}
 \left\langle \mathcal{O}_{\mathbf{u},\mathbf{n}\boldsymbol{\nu}}\right\rangle
 \propto \frac{\mathbbm{C}_{\mathbf{K}}}{\sqrt{J}}\left(\frac{v}{2g}\right)^J \sqrt{
 \frac{\prod\limits_{j}^{}x_{1j}\prod\limits_{j}^{}x_{7j}\prod\limits_{j}^{}\sigma _B(u_{4j})\frac{u_{4j}\left(u_{4j}+\frac{i}{2}\right)}{\sigma(u_{4j},\bar{u}_{4j})} }
 {\prod\limits_{j}^{}x_{3j}\prod\limits_{j}^{}x_{5j}
 \prod\limits_{j}^{}u_{2j}\left(u_{2j}+\frac{i}{2}\right)
 \prod\limits_{j}^{}u_{6j}\left(u_{6j}+\frac{i}{2}\right)}\,
 \,\, \mathop{\mathrm{Sdet}}G
 }.
\end{equation}
Here $J$ is the R-charge of the operator, and $\mathbf{u}$ are rapidities of magnons. The indices $\mathbf{n}$ and $\boldsymbol{\nu}$ label superconformal descendants of the primary operator $\mathcal{O}_{\bf u}$. The superdeterminant $\mathop{\mathrm{Sdet}}G$ is built from a matrix (Gaudin-like matrix \cite{deLeeuw:2015hxa,Kristjansen:2020vbe}) with entries depending on the rapidities, while $\sigma$  and $\sigma_B$ are bulk and boundary scalar factors respectively. See section \ref{sec:bootstrap} for details. A novelty as compared to  the previous works \cite{Komatsu:2020sup,Jiang:2019xdz,Gombor:2020kgu,Gombor:2020auk} is the kinematical factor $\mathbbm{C}_{\mathbf{K}}$, which depends on quantum numbers of the operator and is given by a ratio of Gamma functions. This extra factor is needed since superconformal primary states, which can be built directly from the Bethe ansatz, have vanishing one-point functions and only their specific superconformal descendants have non-zero expectation values. Similar factors appeared also in the analysis of defect one-point functions \cite{Gombor:2024api} and the Separation of Variables approach to the three-point functions \cite{Bercini:2022jxo}. 

We then test the asymptotic formula against direct field theory computations at tree level and one loop, finding perfect agreement. This provides direct evidence for integrability on the Coulomb branch.

The rest of the paper is organized as follows. In {\bf section} \ref{sec:bootstrap}, we determine the boundary reflection matrix for the probe D3-brane by solving integrable boundary bootstrap equations. We then use the result to write down a conjecture for the asymptotic formula for one-point functions.  In {\bf section} \ref{sec:treelevel}, we test the conjecture against the tree-level one-point functions computed on the field-theory side. In particular, we explain how the kinematical factor is reproduced after performing the fermionic duality transformation to the known overlap formulas in the SO(6) spin chain. In {\bf section} \ref{sec:1loop}, we study the one-loop correction and demonstrate that the asymptotic formula correctly reproduces the results from field theory. This provides nontrivial tests for both the boundary dressing phase and the kinematical factor. We conclude and discuss future directions in {\bf section} \ref{sec:conclusion}.
    
    \section{Integrable bootstrap at finite coupling}\label{sec:bootstrap}
    The string sigma model on $AdS_5\times S^{5}$ is integrable \cite{Bena:2003wd} and the worldsheet S-matrix for excitations ({\it magnons}) is known as an exact function of the 't Hooft coupling $\lambda$ \cite{Beisert:2005tm,Beisert:2006qh}. In this section, we leverage this knowledge and conjecture an explicit expression for the vacuum condensates at finite coupling.
    
    The key enabling insight comes from holography: as mentioned in the introduction, vacuum condenstates on the Coulomb branch correspond to overlaps between on-shell closed strings and a boundary state representing a probe D3-brane in AdS. There is supporting evidence---both from weak and strong coupling analyses \cite{Ivanovskiy:2024vel,Dekel:2011ja}---that the corresponding D-brane realizes an {\it integrable boundary}, one that preserves infinitely many conserved charges. 
    
    As discussed in \cite{Ghoshal:1993tm}, integrability imposes strong constraints on such boundaries. In particular, the reflection of excitations at the boundary is governed by a set of consistency relations known as {\it boundary bootstrap equations}. These equations constrain the boundary reflection matrix, which encodes the scattering of magnons against the integrable boundary.
    
    Below, we formulate and solve the boundary bootstrap equations and use the result to write down a conjecture for the asymptotic formula for the vacuum condensate. Our analysis draws on results from the bootstrap of defect one-point functions \cite{Komatsu:2020sup}. We find that the boundary reflection matrix relevant for the Coulomb branch can be obtained as a specific limit of the reflection matrix used in the defect setup. This correspondence allows us to import and adapt those results,  the solution to the crossing equation and the asymptotic formula for vacuum condensates.
    \subsection{Vacuum condensates and spin-chain overlaps} Before discussing the integrable bootstrap, let us first review the results in \cite{Ivanovskiy:2024vel} and recap the relationship between one-point functions on the Coulomb branch and spin-chain overlaps.

    In CFTs, the conformal symmetry forbids local operators to have one-point functions. However, once the theory is on the moduli space and the conformal symmetry is spontaneously broken, they can acquire nonzero expectation values of the following form:
    \beq
\langle \mathcal{O}\rangle_v= c_{\mathcal{O}}v^{\Delta_{\mathcal{O}}}\period
    \eeq
    Here $\Delta_{\mathcal{O}}$ is the conformal dimension of the operator $\mathcal{O}$ and $v$ is the mass scale of the spontaneous symmetry breaking. The constant $c_{\mathcal{O}}$ contains the dynamical information and can be regarded as an analog of the structure constants. In $\mathcal{N}=4$ SYM, $v$ is the expectation value of the scalar field $\phi$ (see \eqref{higgsvev}) and $c_{\mathcal{O}}$ is a function of the coupling constant. 

\begin{figure}[t]
\centering
 \centerline{\includegraphics[clip,height=5cm]{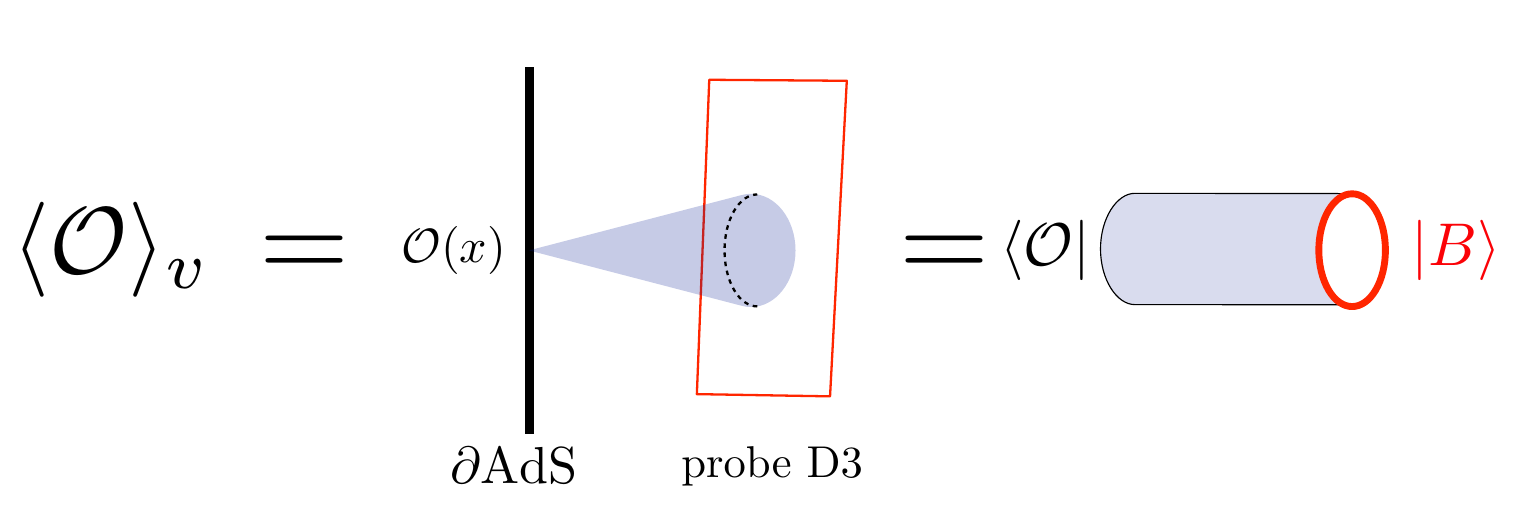}}
\caption{\label{fig:fig1}\small String-theory description of the vacuum condensate. Holography maps the vacuum condensate to a process in which a closed string emitted from the operator $\mathcal{O}(x)$ gets absorbed by a probe D3 brane in $AdS_5\times S^5$. On the worldsheet, this corresponds to an overlap between an on-shell closed string corresponding to $\mathcal{O}(x)$ and a boundary state representing the probe D3 brane.}
\end{figure}
    
    \paragraph{Weak coupling.}In $\mathcal{N}=4$ SYM at tree level, $c_{\mathcal{O}}$ can be evaluated simply by replacing the fields in $\mathcal{O}$ by its classical expectation value \eqref{higgsvev}. Concretely, taking a single-trace operator in the SO(6) sector 
    \beq
    \mathcal{O}=\Psi^{i_1,\ldots, i_{L}}{\rm tr}\left(\phi_{i_1}\cdots\phi_{i_{L}}\right)\comma
    \eeq
    and substituting the fields with their expectation values \eqref{higgsvev}, we obtain
    \beq\label{eq:treeSO6}
\langle \mathcal{O}\rangle_v=v^{L}\Psi^{i_1,\ldots, i_{L}} n_{i_1}\cdots n_{i_{L}}\period
    \eeq
    Here $\Psi^{i_1,\ldots, i_{L}}$ is an eigenfunction of the one-loop dilatation operator and $n_i\equiv v_i/v$. Using the relation between the one-loop dilatation operator and the Hamiltonian of the SO(6) spin chain, \eqref{eq:treeSO6} can be expressed as an overlap in the spin chain\footnote{The prefactor $(8\pi^2 v^2/\lambda)^{L/2}$ and $1/\sqrt{L\langle \Psi|\Psi\rangle}$ come from normalizing the operator $\mathcal{O}$ so that it has a unit two-point function $\langle \mathcal{O}(x)\mathcal{O}(0)\rangle =|x|^{-2\Delta_{\mathcal{O}}}$.},
    \beq
\langle \mathcal{O}\rangle_{v}=\left(\frac{8\pi^2v^2}{\lambda}\right)^{\frac{L}{2}}\frac{\langle B|\Psi\rangle}{\sqrt{L\langle \Psi|\Psi\rangle}}\comma
    \eeq
    where $|\Psi\rangle$ is an eigenstate of the SO(6) spin chain corresponding to the operator $\mathcal{O}$ while $\langle B|$ is
    \beq
\langle B|=n_{i_1}\cdots n_{i_{L}}\langle i_1,\ldots, i_{L}|\period
    \eeq
    Importantly, the state $\langle B|$ belongs to a special class of states called integrable boundary state, which is annihilated by infinitely many conserved charges. Thanks to this property, the overlap $\langle B|\Psi\rangle$ shows two key features
    \begin{itemize}
    \item The overlap $\langle B|\Psi\rangle$ is nonzero if and only if the Bethe state $|\Psi\rangle$ is parity invariant, i.e.~ the rapidities of magnons comes in parity-invariant pairs $|u_1,-u_1,u_2,-u_2\cdots\rangle$.
    \item The overlap admits a closed-form expression in terms of the Gaudin-like determinant \cite{deLeeuw:2015hxa,Kristjansen:2020vbe}.
    \end{itemize}
    
    \begin{figure}[t]
\centering
 \centerline{\includegraphics[clip,height=4cm]{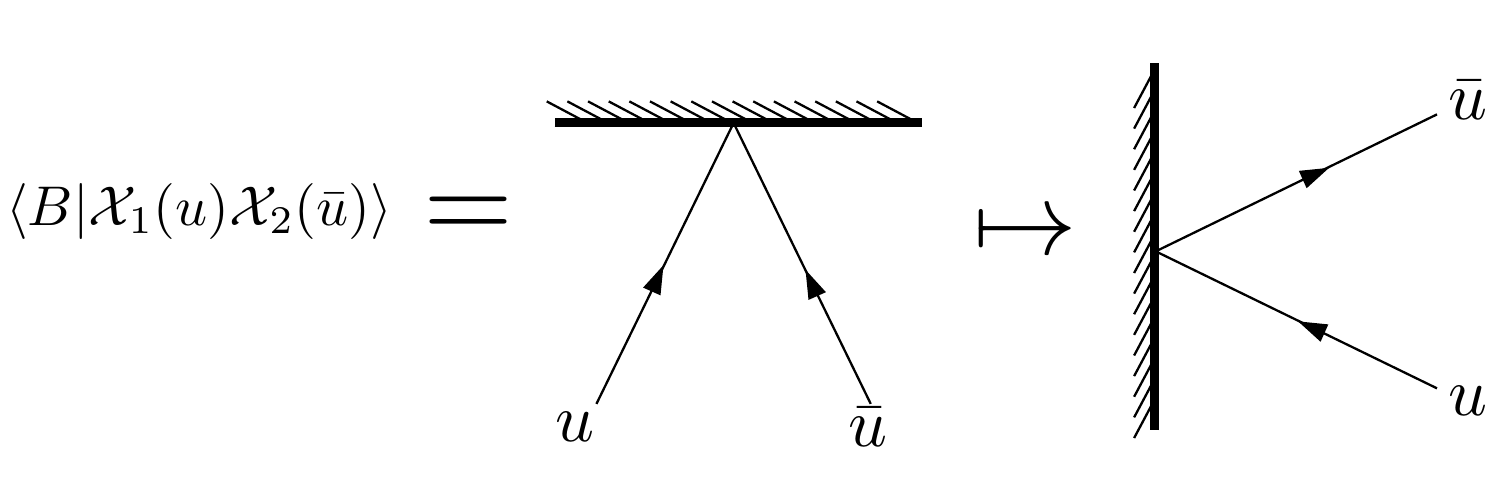}}
\caption{\label{fig:fig2}\small Two-particle form factor and reflection matrix. The two-particle form factor (the left and the middle figures) are related to the reflection matrix (the right figure) by the {\it mirror transformation}, which implements the Wick rotation exchanging space and time.}
\end{figure}

    \paragraph{Finite coupling.} Holography suggests that these remarkable properties persist even at finite coupling. In particular, the interpretation of the vacuum condensates as the overlap still holds at finite coupling: On the string-theory side, the vacuum condensates correspond to overlaps between an on-shell closed string state $|\Psi\rangle$ describing a single-trace operator and the boundary state $|B\rangle$, which describes a probe D-brane in AdS (see Figure \ref{fig:fig1}):
\beq\label{eq:defCoulomb}
\langle \mathcal{O}\rangle_{v}\sim \langle B|\Psi \rangle\period
\eeq
Thanks to integrability, closed string states with large $R$ charge can be described as a collection of ``magnon" excitations as
\beq
|\Psi \rangle = |\mathcal{X}_1(u_1)\mathcal{X}_2(u_2)\cdots \mathcal{X}_M(u_M)\rangle
\eeq
where $\mathcal{X}_i$'s are magnons on top of the BPS states made out of $Z$ fields and $u_i$'s are their rapidities. Thus, the right hand side of \eqref{eq:defCoulomb} is a function of these rapidities and it can be regarded as a kind of form factors.  

Since the boundary state is translationally invariant, the form factor vanishes unless the momenta of magnons add up to zero. This in particular means that the one-particle form factor is nonzero only when the magnon has a zero momentum. More interesting is a two-particle form factor given by
\beq
\langle B|\mathcal{X}_1 (u)\mathcal{X}_2 (\bar{u}) \rangle
\eeq
where $\bar{u}$ is a parity-flipped rapidity (see \eqref{eq:parity} for the definition).
Unlike the one-particle form factor, it is a function of the rapidity $u$. Furthermore, by the Wick rotation, it is related to the reflection matrix of a single magnon at the boundary, see Figure \ref{fig:fig2}. If the boundary is an integrable boundary condition, the single-magnon reflection matrix contains all the information for describing the dynamics of the system. In other words, there is a certain factorization property for the form factors, and knowing the two-particle form factor is enough to determine them all.

In the rest of this section, we determine the structure of the two-particle form factor assuming that the boundary is integrable.

    \subsection{Dynamic $\mathfrak{su}(2|2)$ spin chain}
In order to set the notation and convention, we now review the {\it dynamic $\mathfrak{su}(2|2)$ spin chain} \cite{Beisert:2005tm,Beisert:2006qh}, which describes magnons at finite coupling. 

In the dynamic $\mathfrak{su}(2|2)$ spin chain, magnons belong to the bifundamental representation of $\mathfrak{psu}(2|2)^2$ symmetry. The precise relation between the fields in $\mathcal{N}=4$ SYM and the excitations in the spin chain is given by
\beq
\begin{aligned}
&\phi^{a}\dot{\phi}^{\dot{b}}\mapsto \Phi^{a\dot{b}} &&(\Delta^{0}, J)=(1,0)\\
&\psi^{\alpha}\dot{\psi}^{\dot{\beta}}\mapsto D^{\alpha\dot{\beta}}Z&&(\Delta^{0}, J)=(2,1)\\
&\psi^{\alpha}\dot{\phi}^{\dot{a}}\mapsto  \Psi^{\alpha\dot{a}}&&(\Delta^{0}, J)=(3/2,1/2)\\
&\phi^{a}\dot{\psi}^{\dot{\alpha}}\mapsto \Psi^{a\dot{\alpha}}&&(\Delta^{0}, J)=(3/2,1/2)
\end{aligned}
\eeq
where $\Phi$ and $\Psi$ are the scalar field and the fermion field of $\mathcal{N}=4$ SYM respectively,. $\Delta^{0}$ and $J$ are the bare scaling dimension and the U(1) R-charge respectively\fn{$Z$ and $\bar{Z}$ have $+1$ and $-1$ charges for this U(1) symmetry.}.

Let us now summarize the action of the $\mathfrak{psu}(2|2)^2$ generators on the excitations in the spin chain\fn{In what follows, $\epsilon^{12}=\epsilon^{\dot{1}\dot{2}}=-\epsilon_{12}=-\epsilon_{\dot{1}\dot{2}}=1$.}:
\beq\label{eq:su22sym}
\begin{aligned}
&R^{a}_{b} \ket{\phi^{c}}_{L}= \delta_{b}^{c}\ket{\phi^{a}}_{L}-\frac{1}{2}\delta^{a}_{b}\ket{\phi^{c}}_{L}\comma\quad
&&L^{\alpha}_{\beta}\ket{\psi^{\gamma}}_{L}=\delta^{\gamma}_{\beta} \ket{\psi^{\alpha}}_{L}-\frac{1}{2}\delta^{\alpha}_{\beta}\ket{\psi^{\gamma}}_{L}\comma\\
&Q^{\alpha}_{a}\ket{\phi^{b}}_{L}=a  \delta^{b}_{a}\ket{\psi^{\alpha}}_{L}\comma\quad &&Q^{\alpha}_{a}\ket{\psi^{\beta}}_{L}= b \epsilon^{\alpha\beta}\epsilon_{ab}\ket{Z^{+}\phi^{b}}_{L}\comma\\
&S^{a}_{\alpha}\ket{\phi^{b}}_{L}= c \epsilon^{ab}\epsilon_{\alpha\beta}\ket{Z^{-}\psi^{\beta}}_{L}\comma\quad &&S^{a}_{\alpha}\ket{\psi^{\beta}}_{L}=d \delta^{\beta}_{\alpha}\ket{\phi^{a}}_{L}\period 
\end{aligned}
\eeq
\beq
\begin{aligned}
&\dot{R}^{\dot{a}}_{\dot{b}} \ket{\phi^{\dot{c}}}_{R}= \delta_{\dot{b}}^{\dot{c}}\ket{\phi^{\dot{a}}}_{R}-\frac{1}{2}\delta^{\dot{a}}_{\dot{b}}\ket{\phi^{\dot{c}}}_{R}\comma\quad
&&\dot{L}^{\alpha}_{\dot{\beta}}\ket{\psi^{\dot{\gamma}}}_{R}=\delta^{\dot{\gamma}}_{\dot{\beta}} \ket{\psi^{\dot{\alpha}}}_{R}-\frac{1}{2}\delta^{\dot{\alpha}}_{\dot{\beta}}\ket{\psi^{\dot{\gamma}}}_{R}\comma\\
&\dot{Q}^{\dot{\alpha}}_{\dot{a}}\ket{\phi^{\dot{b}}}_{R}=\dot{a}  \delta^{\dot{b}}_{\dot{a}}\ket{\psi^{\dot{\alpha}}}_{R}\comma\quad &&\dot{Q}^{\dot{\alpha}}_{\dot{a}}\ket{\psi^{\dot{\beta}}}_{R}= \dot{b} \epsilon^{\dot{\alpha}\dot{\beta}}\epsilon_{\dot{a}\dot{b}}\ket{Z^{+}\phi^{\dot{b}}}_{R}\comma\\
&\dot{S}^{\dot{a}}_{\dot{\alpha}}\ket{\phi^{\dot{b}}}_{R}= \dot{c} \epsilon^{\dot{a}\dot{
b}}\epsilon_{\dot{\alpha}\dot{\beta}}\ket{Z^{-}\psi^{\dot{\beta}}}_{R}\comma\quad &&\dot{S}^{\dot{a}}_{\dot{\alpha}}\ket{\psi^{\dot{\beta}}}_{R}=\dot{d} \delta^{\dot{\beta}}_{\dot{\alpha}}\ket{\phi^{\dot{a}}}_{R}\period 
\end{aligned}
\eeq
\beq
\begin{aligned}
&C\ket{\mathcal{X}}= \frac{1}{2}(ad+bc)\ket{\mathcal{X}}\comma\quad &&\dot{C}\ket{\dot{\mathcal{X}}}=\frac{1}{2}(\dot{a}\dot{d}+\dot{b}\dot{c})\ket{\dot{\mathcal{X}}}\comma\\
&P\ket{\mathcal{X}}=ab\ket{Z\mathcal{X}}\comma\quad &&\dot{P}\ket{\dot{\mathcal{X}}}=\dot{a}\dot{b}\ket{Z\dot{\mathcal{X}}}\comma\\
&K\ket{\mathcal{X}}=cd\ket{Z^{-}\mathcal{X}}\comma\quad && \dot{K}\ket{\dot{\mathcal{X}}}=\dot{c}\dot{d}\ket{Z^{-}\dot{\mathcal{X}}}\comma\\
&\ket{\mathcal{X}Z^{\pm}}=\left( \frac{x^{+}}{x^{-}}\right)^{\pm1}\ket{Z^{\pm}\mathcal{X}}\comma\quad &&\ket{\dot{\mathcal{X}}Z^{\pm}}=\left( \frac{x^{+}}{x^{-}}\right)^{\pm1}\ket{Z^{\pm}\dot{\mathcal{X}}}\comma
\end{aligned}
\eeq
Here $L$'s are Lorentz generators and $R$'s are the R-symmetry generators while $Q$ and $S$ are the supersymmetry and the superconformal generators respectively. $C$, $P$ and $K$ are the central charges of $\mathfrak{psu}(2|2)^2$, where $C$ is related to the anomalous dimension
\beq
C=\frac{D-J}{2}\comma
\eeq
and $P$ and $K$ are the field-dependent gauge transformations\fn{ A field-theoretic interpretation of this algebra and the representation is given in section 3.2 in \cite{Komatsu:2017buu}.}.

The parameters $a$-$d$ are given by
\beq
a=\dot{a}=\sqrt{g}\gamma\comma \quad b=\dot{b}= \frac{\sqrt{g}}{\gamma}\left( 1-\frac{x^{+}}{x^{-}}\right)\comma \quad c=\dot{c}=\frac{i\sqrt{g} \gamma}{ x^{+}}\comma \quad d =\dot{d}=\frac{ \sqrt{g}x^{+}}{i\gamma}\left( 1- \frac{x^{-}}{x^{+}}\right)\comma
\eeq
where $\gamma$ satisfies
\beq
|\gamma|^2 =  i(x^{-}-x^{+}) \period
\eeq
and $x(u)$ is the Zhukowsky variable given by
\beq\label{zhuk}
u=g \left(x+\frac{1}{x} \right)\period
\eeq
The plus and minus superscripts denote the shift of the rapidity by $i/2$, namely $f^{\pm}(u)=f(u\pm i/2)$. In terms of the Zhukowsky variables, the momentum and the energy of a magnon is given by
\beq
E(u)=\frac{1}{2} \frac{1+\frac{1}{x^{+}x^{-}}}{1-\frac{1}{x^{+}x^{-}}}\comma\qquad e^{ip(u)}=\frac{x^{+}}{x^{-}}
\eeq
The total momentum of an on-shell state should be zero,  while the total energy determines  the scaling dimension of the gauge-theory operator: 
\begin{equation}\label{totalE}
 \Delta =J+2\sum_{j}^{}E(u_j).
\end{equation}

A couple of comments are in order: First, the $Z$-markers appearing on the right hand sides of the transformation rules are neccessary in order to match the $\Delta^{0}$ and $J$ charges on both sides. 
Second, although there are two copies of $\mathfrak{psu}(2|2)$, there is only one set of central charges $P,K$ and $C$. Third, the parity-inverted rapidity is defined by\fn{$\bar{u}$ is basically $-u$ but since $x^{\pm}$ are multi-valued functions of $u$, we defined it in this way.}
\beq\label{eq:parity}
x^{+}(\bar{u})=-x^{-}(u)\comma\qquad x^{-}(\bar{u})=-x^{+}(u)\period
\eeq

\subsection{Bootstrapping form factor I: symmetry constraints}
We now start our main task of determining the two-particle form factor $\langle B|\mathcal{X}_1 (u)\mathcal{X}_2 (\bar{u}) \rangle$. The first constraints come from symmetry. As explained above, the spin chain for $\mathcal{N}=4$ SYM has the centrally-extended $\mathfrak{psu}(2|2)^2$ as its symmetry group. If we give an expectation value to a scalar contained in $Z$, we lose all the superconformal symmetry but retain all the supersymmetry and the bosonic subgroups of $\mathfrak{psu}(2|2)^2$. This means that the boundary state $\langle B |$ is annihilated by those unbroken symmetry generators $J$
\beq
\langle B| J=0\period
\eeq 
In what follows, we impose
\beq
\langle B| J|\mathcal{X}_1 (u)\mathcal{X}_2 (\bar{u})\rangle= 0\comma
\eeq
and determine the structure of the two-particle form factor. Since the symmetry group has a product structure ($\mathfrak{psu}(2|2)^2$) and each magnon belongs to a bifundamental representation, the form factor factorizes into contributions from the left and right $\mathfrak{psu}(2|2)$'s. Thus below we discuss each $\mathfrak{psu}(2|2)$ separately.
\paragraph{Constraints from bosonic symmetries.} For two particles, the only invariant of the bosonic symmetries are $\epsilon$ tensors. We thus conclude that the form factor takes the following form in order for it to be compatible with the bosonic symmetries:
\beq\label{eq:symconstraints}
\begin{aligned}
&\langle B | \phi^{a}(u) \phi^{b}(\bar{u})\rangle =k_0 (u)\epsilon^{ab}\comma\qquad \langle B | \psi^{\alpha}(u) \psi^{\beta}(\bar{u})\rangle =k_1 (u)\epsilon^{\alpha\beta}\\
&\text{others} =0\period
\end{aligned}
\eeq
\paragraph{Constraints from supersymmetries.}
Let us next consider the constraints from the supersymmetries. In particular we impose the condition
\beq
\langle B | Q^{\alpha}{}_{a}|\psi^{\beta}(u)\phi^{b}(\bar{u})\rangle=0\period
\eeq
Acting $Q$ on the ket, using \eqref{eq:su22sym}, we get
\beq
\begin{aligned}
\langle B | Q^{\alpha}{}_{a}|\psi^{\beta}(u)\phi^{b}(\bar{u})\rangle&=b(u)\epsilon^{\alpha\beta}\epsilon_{ac}\langle B|Z\phi^{c}\phi^{b}\rangle-a(\bar{u})\delta^{b}_{a}\langle B|\psi^{\beta}\psi^{\alpha} \rangle \\
&=\epsilon^{\alpha\beta}\delta_{a}^{b}\left(\kappa b(u)k_1 (u)+a(\bar{u})k_0 (u)\right)\period 
\end{aligned}
\eeq
Here we {\it assumed} that an extra insertion of $Z$ costs $\kappa$:
\beq
\langle B | Z \Psi \rangle =\kappa \langle B |\Psi \rangle\period
\eeq
Physically this parameter $\kappa$ is related to the coordinate of the Coulomb branch moduli $v$. By comparing the final result with the one-point function of the chiral primary, we can show $\kappa$ is related to $v$ by
\beq
\kappa =\frac{v}{2g}\period
\eeq

Imposing the symmetry constraint, we have
\beq
\frac{k_1 (u)}{k_0 (u)}=-\frac{a(\bar{u})}{\kappa b(u)}= -\frac{x^{-}\gamma (\bar{u})\gamma (u)}{\kappa(x^{-}-x^{+})}=\frac{x^{-}}{i\kappa}
\eeq

One can also impose another constraint,
\beq
\begin{aligned}
\langle B | Q^{\alpha}{}_{a}|\phi^{b}(u)\psi^{\beta}(\bar{u})\rangle= 0\period
\end{aligned}
\eeq
Computing the left hand side by acting the generator to the ket, we get
\beq
\begin{aligned}
\langle B | Q^{\alpha}{}_{a}|\phi^{b}(u)\psi^{\beta}(\bar{u})\rangle&=a(u)\delta_{a}^{b}\langle B | \psi^{\alpha}\psi^{\beta}\rangle+b(\bar{u})\epsilon^{\alpha\beta}\epsilon_{ac}\langle B |\phi^{b}Z \phi^{c}\rangle\comma\\
&=\epsilon^{\alpha\beta}\delta^{a}_{b}\left(a(u) k_0 (u)-\kappa b(\bar{u}) e^{ip} k_1 (u) \right)\period
\end{aligned}
\eeq
Thus $k_1/k_0$ must satisfy
\beq
\begin{aligned}
\frac{k_1 (u)}{k_0 (u)}=\frac{a (u)}{\kappa b(\bar{u})e^{ip}}=-\frac{x^{-}\gamma (\bar{u})\gamma (u)}{\kappa (x^{-}-x^{+})}=\frac{x^{-}}{i\kappa}\period
\end{aligned}
\eeq
As expected, this reproduces the previous result and serves as a consistency check.
\paragraph{Summary.}
To summarize, the symmetry constraints completely determines the matrix structure of the two-particle form factor and the only undetermined piece is the overall scalar factor $k (u)$:
\beq
\begin{aligned}
&\langle B |\mathcal{X}^{A\dot{A}}(u)\mathcal{X}^{B\dot{B}}(\bar{u})\rangle =k (u) \mathcal{C}^{AB}\mathcal{C}^{\dot{A}\dot{B}}
\end{aligned}
\eeq
(Note that here we redefined $k_0(u)^2$ in the previous subsection by $k(u)$.)
The matrix-parts $\mathcal{C}^{AB}$ and $\mathcal{C}^{\dot{A}\dot{B}}$ are given by the structure we determined above:
\beq
\begin{aligned}
\mathcal{C}^{AB}:\quad &\langle B | \psi^{\alpha}(u)\psi^{\beta}(\bar{u})\rangle = \epsilon^{\alpha\beta}\comma\\
&\langle B | \phi^{a}(u)\phi^{b}(\bar{u})\rangle = \frac{x^{-}}{i\kappa}\epsilon^{ab}\period
\end{aligned}
\eeq
\subsection{Bootstrapping form factor II: scalar factor}
 Having determined the matrix structure of the two-particle form factor, the next task is to constrain the overall scalar factor. To do this, we impose two different constraints, both of which are often called the form factor axioms.
\paragraph{String frame.}
To study the form factor axioms, it is customary to change the definition of the $\phi$ excitation as
\beq
\phi_{\rm new}=Z^{1/4} \phi_{\rm old}Z^{1/4}\period
\eeq 
The excitations defined in this way are called {\it string-frame} excitations while the original ones are called {\it spin-chain frame} excitations\fn{The original normalization is natural for the comparison with the gauge theory while the new one is more natural from the world-sheet point of view.}.

\begin{figure}[t]
\centering
\begin{minipage}{0.48\hsize}
\centering
 \includegraphics[clip,height=2.5cm]{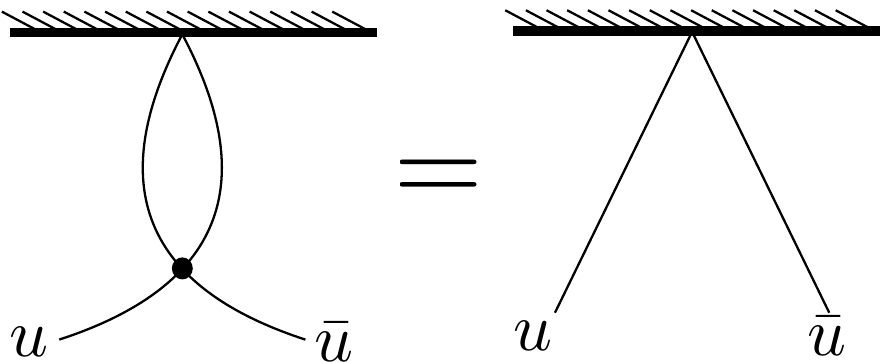}\\
 \text{1.~Watson equation}
 \end{minipage}
 \begin{minipage}{0.48\hsize}
\centering
 \includegraphics[clip,height=2.5cm]{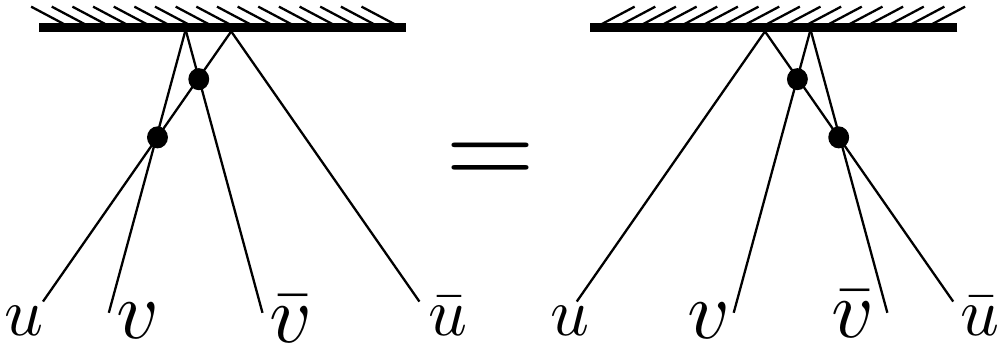}\\
 \text{2.~Boundary Yang-Baxter equation}
 \end{minipage}
\caption{\label{fig:WatsonbYB}\small Watson equation and boundary Yang-Baxter equation. The black dots represent the action of the S-matrix.}
\end{figure}

After this change, the two-particle form factor for the scalar gets modified as follows:
\beq
\begin{aligned}
\left.\langle B | \phi^{a}(u)\phi^{b}(\bar{u})\rangle \right|_{\rm string}&= \left.\langle B |Z^{1/4}\phi^{a}(u)Z^{1/2}\phi^{b}(\bar{u})Z^{1/4}\rangle\right|_{\rm spin}\\
&=\kappa e^{ip/2}\left.\langle B |\phi^{a}(u)\phi^{b}(\bar{u})\rangle\right|_{\rm spin}
\end{aligned}
\eeq
Therefore, the two-particle form factors in the string frame read
\beq
\begin{aligned}\label{eq:matrixstring}
\mathcal{C}^{AB}_{\rm string}:\quad&\langle B | \psi^{\alpha}(u)\psi^{\beta}(\bar{u})\rangle = \epsilon^{\alpha\beta}\comma\\
&\langle B | \phi^{a}(u)\phi^{b}(\bar{u})\rangle = \frac{\sqrt{x^{+}x^{-}}}{i}\epsilon^{ab}\period
\end{aligned}
\eeq
\paragraph{Watson equation.}
The first constraint is called the Watson equation, see Figure \ref{fig:WatsonbYB}. It essentially tells us that if we swap the order of the particles, we need to multiply the corresponding S-matrix. Written explicitly, it says
\beq
\langle B | \mathbb{S} |\mathcal{X}_1 \mathcal{X}_2\rangle =\langle B |\mathcal{X}_1 \mathcal{X}_2\rangle\comma
\eeq
where $\mathbb{S}$ is the two-particle S-matrix. One can show (using Mathematica) that this boils down to the following equation for the overall factor $k(u)$:
\beq
\frac{k (u)}{k(\bar{u})}=\left(\frac{x^{+}}{x^{-}}\right)^2 S_0 (u,\bar{u})\period
\eeq
Here $S_0 (u,\bar{u})$ is the scalar part of the $\mathfrak{psu}(2|2)$ S-matrix given by
\beq
S_0 (u,\bar{u})=\frac{x^{+}}{x^{-}}\frac{1+1/(x^{-})^2}{1+1/(x^{+})^2}\frac{1}{\sigma^2 (u,\bar{u})}\comma
\eeq
where $\sigma(u,v)$ is the dressing phase \cite{Beisert:2006ez}.
In summary, the Watson equation is given by
\beq
\frac{k (u)}{k(\bar{u})}=\left(\frac{x^{+}}{x^{-}}\right)^3\frac{1+1/(x^{-})^2}{1+1/(x^{+})^2}\frac{1}{\sigma^2 (u,\bar{u})}\period
\eeq

\begin{figure}[t]
\centering
 \centerline{\includegraphics[clip,height=3cm]{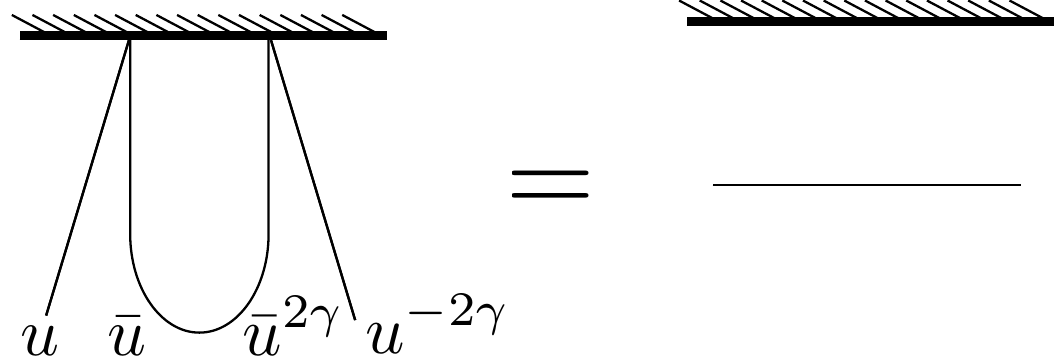}}
\caption{\label{fig:crossing}\small The crossing equation for the two-particle form factor. It amounts to requiring that the form factor of a pair of singlets (the one formed by $u$ and $u^{-2\gamma}$ and the one formed by $\bar{u}$ and $\bar{u}^{2\gamma}$) is trivial. In general, one needs to sum over all possible states forming a singlet pair. However, in the current case, it simplifies to \eqref{eq:crossing} by judiciously choosing the external states.}
\end{figure}

\paragraph{Boundary Yang-Baxter equation.} The second constraint is the boundary Yang-Baxter equation shown in Figure \ref{fig:WatsonbYB}.
Using the matrix part we determined in the previous section, one can check that the boundary Yang-Baxter equation is satisfied with our solution, irrespective of the overall factor $k(u)$. This provides strong evidence for the integrability of the boundary state.
\paragraph{Crossing equation.}
The last constraint is the so-called crossing equation. It requires the form factor to be trivial when magnons form singlets (see Figure \ref{fig:crossing}). In our case, it boils down to the following relation\fn{To see that the rapidities $\bar{u}^{2\gamma}$ and $u^{-2\gamma}$ are related by the parity, we need to use the relation $\overline{u^{2\gamma}}=\bar{u}^{-2\gamma}$.
}:
\beq\label{eq:crossing}
\langle B|\underbrace{D^{1\dot{1}} (u)D^{2\dot{2}}(\bar{u})}_{\rm singlet}\underbrace{D^{1\dot{1}}(\bar{u}^{2\gamma})D^{2\dot{2}}(u^{-2\gamma})}_{\rm singlet}\rangle=1\period
\eeq
Here $u\to u^{2\gamma}$ is the {\it crossing transformation}. For details about the crossing transformation in the context of overlaps with a boundary state, see \cite{Jiang:2019zig}.
To compute the left hand side, one has to analyze the four-particle form factor. Here we assume that the system is described by an integrable boundary state and the four-particle form factor is given by a product of two two-particle form factors.  Under this assumption, \eqref{eq:crossing} becomes
\beq
k(u)k(\bar{u}^{2\gamma})=1\period
\eeq

\subsection{Solving the Watson and crossing equations}
\paragraph{Simplifying the equations.}We now determine the scalar factor $k(u)$ by solving the Watson and crossing equations. For this purpose, we first rewrite $k(u)$ as
\beq
k(u)=\frac{x^{+}+x^{-}}{2(x^{-})^3}\frac{x^{-}+\frac{1}{x^{-}}}{x^{-}+\frac{1}{x^{+}}}\frac{\sigma_B(u)}{\sigma (u,\bar{u})}\period
\eeq
Here $\sigma_B$ is the boundary dressing phase which is yet to be determined. With this rewriting, the Watson equation becomes the following simple relation for $\sigma_B$:
\beq
\sigma_B(\bar{u})=\sigma_B (u)\period
\eeq
%%%
On the other hand, the crossing equation imposes a somewhat nontrivial constraint on the dressing phase $\sigma_B(u)$:
\beq
\begin{aligned}
&1=\frac{x^{+}+x^{-}}{2(x^{-})^3}\frac{x^{-}+\frac{1}{x^{-}}}{x^{-}+\frac{1}{x^{+}}}\frac{\sigma_B(u)}{\sigma (u,\bar{u})}\frac{(x^{+})^2(x^{+}+x^{-})(1+(x^{+})^2)}{2x^{-}(1+x^{+}x^{-})}\frac{\sigma_B(\bar{u}^{2\gamma})}{\sigma (\bar{u}^{2\gamma},u^{-2\gamma})}\\
&\iff \sigma_B(u)\sigma_B(\bar{u}^{2\gamma})=\frac{4 (x^{-})^{4}}{(x^{+}+1/x^{+})(x^{-}+1/x^{-})}\left(\frac{1+\frac{1}{x^{+}x^{-}}}{x^{+}+x^{-}}\right)^2\left(\frac{x^{-}}{x^{+}}\right)^2\sigma (u,\bar{u})\sigma (\bar{u}^{2\gamma},u^{-2\gamma})\period
\end{aligned}
\eeq
To compute the product of the dressing phases $\sigma (u,\bar{u})\sigma (\bar{u}^{2\gamma},u^{-2\gamma})$, we rewrite it as
\beq
\sigma (u,\bar{u})\sigma (\bar{u}^{2\gamma},u^{-2\gamma})=\sigma (u,\bar{u})\sigma (u^{-2\gamma},\bar{u})\sigma (\bar{u},u^{-2\gamma})\sigma (\bar{u}^{2\gamma},u^{-2\gamma})\comma
\eeq
and use the crossing equation for the dressing phase
\beq
\sigma (u,v)\sigma (u^{2\gamma},v)=\frac{(1-1/x^{+}y^{+})(1-x^{-}/y^{+})}{(1-1/x^{+}y^{-})(1-x^{-}/y^{-})}\period
\eeq
We then get
\beq
\begin{aligned}
\frac{1}{\sigma (u,\bar{u})\sigma (\bar{u}^{2\gamma},u^{-2\gamma})}=&\frac{4 (x^{-})^{4}}{(x^{+}+1/x^{+})(x^{-}+1/x^{-})}\left(\frac{1+\frac{1}{x^{+}x^{-}}}{x^{+}+x^{-}}\right)^2\period
\end{aligned}
\eeq

We therefore obtain the following crossing equation for the boundary dressing phase:
\beq
\sigma_B (u)\sigma_B (\bar{u}^{2\gamma})=\left(\frac{x^{-}}{x^{+}}\right)^2\period
\eeq
Using the parity transformation of the rapidity $\bar{u}^{2\gamma}=\overline{u^{-2\gamma}}$ and the parity invariance of $\sigma_B$, $\sigma_B(u)=\sigma_B (\bar{u})$, we can rewrite this into
\beq\label{eq:crossingtosolve}
\sigma_B(u)\sigma_B (u^{2\gamma})=\left(\frac{x^{+}}{x^{-}}\right)^2\period
\eeq

\paragraph{Solution.}The remaining task is to solve the equation \eqref{eq:crossingtosolve}. This turns out to be staightforward since we can recycle the results for the defect one-point function in \cite{Komatsu:2020sup}. 
Let us first recall the crossing equation for the defect one-point function
\beq\label{eq:defectcrossing}
\sigma_{B}^{D}(u)\sigma_B^{D}(u^{2\gamma})=\frac{1}{x_s^{4}}\frac{(x^{+}-1/x_s)^2(x^{-}+x_s)^2}{(x^{+}+x_s)^2(x^{-1}-1/x_s)^2}\comma
\eeq
where $\sigma_{B}^{D}$ is the boundary dressing phase for the defect one-point function, and $x_s$ is a c-number which depends on the detail of the defect\fn{More precisely, the D5 defect discussed in \cite{Komatsu:2020sup} is labelled by an integer $s$, which is a size of the SU(2) representation entering in the solution to the Nahm pole equation. $x_s$ in the formula is related to this integer $s$ by $is=g(x_s+1/x_s)$. Therefore, the limit $x_s\to\infty$ corresponds to the limit in which the integer $s$ is sent to infinity.}. Comparing \eqref{eq:crossingtosolve} and \eqref{eq:defectcrossing}, one  immediately notices that the solution to \eqref{eq:crossingtosolve} is given formally by 
\beq\label{eq:limittotake}
\hat{\sigma}_B=\lim_{x_s\to\infty}x_s^{2} \sigma_{B}^{D}(u)\period
\eeq
It turns out that this naive limit is divergent and one needs to make use of the CDD factor,
\beq
\sigma_{\rm CDD}(u)=-\left(\frac{v}{-i x_s g}\right)^{4E(u)}\comma
\eeq
in order to obtain a finite result. Relegating these technical details to appendix \ref{ap:sigmaB}, here we present the final result:
\beq\label{sigmaB-final}
\sigma_B(u)=\lim_{x_s\to\infty} x_s^2\sigma_{\rm CDD}(u)\sigma_{B}^{D}(u)=\frac{(v/2)^{4E(u)}}{g^2}e^{i(\tilde{\chi}(x^{+})-\tilde{\chi}(x^{-}))} \comma
\eeq
with
\beq
\tilde{\chi}(x)\equiv-\frac{2}{i}\oint_{|z|=1}\frac{dz}{2\pi i}\oint_{|w|=1}\frac{dw}{2\pi i}\frac{1}{x-z}\frac{1}{w}\log \mathfrak{G}(z,w)\comma
\eeq
and
\beq\label{Gamma-functions}
\mathfrak{G}(z,w)=\frac{\Gamma \left[1+ig (z+\tfrac{1}{z}+w+\tfrac{1}{w})\right]}{\Gamma \left[1-ig (z+\tfrac{1}{z}-w-\tfrac{1}{w})\right]}\period
\eeq
One can explicitly check that the phase (\ref{sigmaB-final}) solves the crossing equation (\ref{eq:crossingtosolve}) as it should. The  technical details of  the derivation are left to the appendix~\ref{appendix:solve-srossing}.

\paragraph{Weak coupling expansion.} Since our solution is the limit of the solution in \cite{Komatsu:2020sup}, we can also read off the weak-coupling expansion from the results in \cite{Komatsu:2020sup}. Up to one loop, we have 
\beq\label{weak-dressing}
\sigma_B(u)=\left(\frac{v}{2g}\right)^2\left(1+\frac{4g^2}{u^2+\frac{1}{4}}\left(\gamma_{E}+\log (v/2)\right)+O(g^4)\right)\period
\eeq
\subsection{Asymptotic one-point function}
Using the result in the previous subsection, we now write down a conjecture for the asymptotic one-point function. As was the case for solving the crossing equation, we make use of the relationship to the defect one-point function, in particular the asymptotic one-point function written down in \cite{Gombor:2020kgu,Gombor:2020auk}, and tested against field-theory computations in \cite{Komatsu:2020sup}.

\paragraph{Conjecture for highest weight states.}A crucial observation is that the matrix part of the form factor \eqref{eq:matrixstring} also coincides (up to an overall factor) with the $x_s\to \infty$ limit of the one for the defect one-point function\fn{This can be checked by comparing \eqref{eq:matrixstring} and (4.30) of \cite{Komatsu:2020sup}.}. This allows us to write down the asymptotic formula by taking the limit of the result in \cite{Gombor:2020kgu,Gombor:2020auk} (see appendix \ref{ap:asym} for details). In the string frame, the result reads
\beq
\begin{aligned}\label{exact-1pt0}
\langle \mathcal{O}\rangle_v=\frac{1}{\sqrt{J}}\left(\frac{v}{2g}\right)^{J}\sqrt{
 \frac{\prod\limits_{j=1}^{K_1}x_{1j}\prod\limits_{j=1}^{K_7}x_{7j}\prod\limits_{j=1}^{K_4}\sigma _B(u_{4j})\frac{u_{4j}\left(u_{4j}+\frac{i}{2}\right)}{\sigma(u_{4j},\bar{u}_{4j})} }
 {\prod\limits_{j=1}^{K_3}x_{3j}\prod\limits_{j=1}^{K_5}x_{5j}
 \prod\limits_{j=1}^{K_2}u_{2j}\left(u_{2j}+\frac{i}{2}\right)
 \prod\limits_{j=1}^{K_6}u_{6j}\left(u_{6j}+\frac{i}{2}\right)}\,
 \,\, \mathop{\mathrm{Sdet}}G_{\rm str}}\period
\end{aligned}
\eeq
Here,
\begin{itemize}
\item $u_{mj}$ denotes the rapidity on the $m$-th node of the Dynkin diagram ($u_{4j}$ is the standard momentum carrying rapidity)and $x_{mj}$ is defined by $g(x_{mj}+1/x_{mj})=u_{mj}$. At each node, the Bethe roots come in parity pairs; $(u_{m1},\bar{u}_{m1},\ldots)$. 
\item ${\rm Sdet}\,G_{\rm str}$ is a super Gaudin determinant written in terms of the Bethe equations in the string frame.
\item The result is written in the ``alternate SU(2) grading", which corresponds to choosing $\eta_1=\eta_2=+1$ in \cite{Beisert:2005fw}. In this grading, operators in the SU(2) sector correspond to states with only $u_{4j}$'s.
\item In order to have a non-zero vacuum condensate, the operator needs to be singlet under the SO(5) R-symmetry and the SO(3,1) Lorentz symmetry. This forces the number of roots to satisfy 
\beq\label{eq:singletcondition}
K_4=2K_2=2K_6=K_1+K_3=K_5+K_7\period
\eeq
 \item Because of (\ref{totalE}), the boundary dressing phase (\ref{sigmaB-final}) promotes $v^J$ to $v^\Delta $ in accord with the dimensional counting, and also shifts the power of the coupling in the prefactor to $L=J+K_4$, the spin-chain length. This is again expected from operator normalization: properly normalized one-point function is divided by the square root of the two-point correlator which scales as $\left\langle \mathcal{O}\,\mathcal{O}\right\rangle\sim g^{2L}$.
\end{itemize}
For the comparison with the field-theory computations, it is more convenient to express it in the spin-chain frame,
\beq
\begin{aligned}\label{exact-1pt}
\langle \mathcal{O}\rangle_v=\frac{1}{\sqrt{L}}\left(\frac{v}{2g}\right)^{J}\sqrt{
 \frac{\prod\limits_{j=1}^{K_1}x_{1j}\prod\limits_{j=1}^{K_7}x_{7j}\prod\limits_{j=1}^{K_4}\sigma _B(u_{4j})\frac{u_{4j}\left(u_{4j}+\frac{i}{2}\right)}{\sigma(u_{4j},\bar{u}_{4j})} }
 {\prod\limits_{j=1}^{K_3}x_{3j}\prod\limits_{j=1}^{K_5}x_{5j}
 \prod\limits_{j=1}^{K_2}u_{2j}\left(u_{2j}+\frac{i}{2}\right)
 \prod\limits_{j=1}^{K_6}u_{6j}\left(u_{6j}+\frac{i}{2}\right)}\,
 \,\, \mathop{\mathrm{Sdet}}G}\comma
\end{aligned}
\eeq
where ${\rm Sdet}\, G$ is a super Gaudin determinant in the spin-chain frame explained below (see \eqref{eq:SdetG}) and the spin-chain length $L$ is given by
\beq
J=L-K_4+\frac{K_3+K_5-K_1-K_7}{2}\period
\eeq
We conjecture \eqref{exact-1pt0} and \eqref{exact-1pt} applies to highest weight states, namely states with no roots at infinity. This is supported by the fact that the asymptotic formula for the defect one-point functions---from which \eqref{exact-1pt0} and \eqref{exact-1pt} are derived---are reasonably well-tested for highest-weight states. However, for the comparison with weak-coupling results, we need to study superconfomal descendants that have Bethe roots at infinity, in order to satisfy the singlet condition \eqref{eq:singletcondition}. The overlap formula is expected to be modified for such states, and we describe the modification below.
\paragraph{Conjecture for descendants.} To discuss descendants, it is convenient to split the Bethe roots into those at infinity $(K_j^{\infty})$ and those that are finite $(\bar{K}_j)$:
\beq
K_{j}=K_j^{\infty}+\bar{K}_j\period
\eeq
We then conjecture that the asymptotic formula (in the spin-chain frame) is given by the following expression:
\beq
\begin{aligned}\label{exact-1pt2}
&\langle \mathcal{O}\rangle_v=\\
&\frac{g^{\frac{K_3^{\infty}+K_5^{\infty}-K_1^{\infty}-K_7^{\infty}}{2}}\mathbbm{C}_{\mathbf{K}}}{\sqrt{L}}\left(\frac{v}{2g}\right)^{J}\sqrt{
 \frac{\prod\limits_{j=1}^{\bar{K}_1}x_{1j}\prod\limits_{j=1}^{\bar{K}_7}x_{7j}\prod\limits_{j=1}^{\bar{K}_4}\sigma _B(u_{4j})\frac{u_{4j}\left(u_{4j}+\frac{i}{2}\right)}{\sigma(u_{4j},\bar{u}_{4j})} }
 {\prod\limits_{j=1}^{\bar{K}_3}x_{3j}\prod\limits_{j=1}^{\bar{K}_5}x_{5j}
 \prod\limits_{j=1}^{\bar{K}_2}u_{2j}\left(u_{2j}+\frac{i}{2}\right)
 \prod\limits_{j=1}^{\bar{K}_6}u_{6j}\left(u_{6j}+\frac{i}{2}\right)}\,
 \,\, \mathop{\mathrm{Sdet}}G}\comma
\end{aligned}
\eeq
\begin{itemize}
\item ${\rm Sdet}$ is a super Gaudin determinant consisting only of finite Bethe roots.
\item $\mathbbm{C}_{\boldsymbol{K}}$ %$\mathbbm{C}_{\mathbf{n}\boldsymbol{\nu}}$ 
is a kinematical prefactor that depends on the quantum number of the state. At tree level, it can be derived by performing the fermionic duality as we explain in section \ref{subsec:duality}. Conjectures for the finite-coupling generalization is presented in section \ref{subsec:descendant}.
\item The prefactor $g^{\frac{K_3^{\infty}+K_5^{\infty}-K_1^{\infty}-K_7^{\infty}}{2}}$ comes from the $u_{mj}\to \infty$ behavior of the relevant factors, $x_{mj}\to  u_{mj}/g$.
\end{itemize}
In the rest of this subsection, we give details on the Bethe equation and the Gaudin determinant in order to clarify the convention.

\paragraph{Bethe equation in the spin-chain frame.} In order to state the Bethe equations we first introduce the auxiliary functions $\chi_a$, with $a=1,\cdots,7$, associated to each node in the Dynkin diagram of the $SU(2)$ grading: 
\beq\label{eq:su2BetheEq}
\def\a{1.5}
\begin{tikzpicture}[baseline={(0,\a)}]  
    \tikzstyle{boson}  = [circle, minimum width=8pt, draw, inner sep=0pt]
    \tikzstyle{fermion}   = [circle, minimum width=8pt, draw, inner sep=0pt, path picture={\draw (path picture bounding box.south east) -- (path picture bounding box.north west) (path picture bounding box.south west) -- (path picture bounding box.north east);}]
%bosonic nodes
\foreach \x/\y/\z in {0/0/4,0/2/2}
{\node[boson,thick,scale=2](\z) at (\x*\a,\y*\a){};}
%fermionic nodes
\foreach \x/\y/\z in {0/1/3,0/3/1}
{\node[fermion,thick,scale=2](\z) at (\x*\a,\y*\a){};}
\foreach \i/\j in {4/3,3/2,2/1}
{\draw[-,thick] (\i) to (\j);}
\node[](v5) at (0,-1*\a){};
\draw[-,dashed,thick] (4) to (v5);
\end{tikzpicture} \quad
\begin{aligned}
e^{i\chi_{1}(z)}=&\prod_{j=1}^{K_2}\frac{z-u_{2,j}+\frac{i}{2}}{z-u_{2,j}-\frac{i}{2}}\prod_{j=1}^{K_4}\frac{1-\frac{1}{x(z)\,x_{4,j}^{+}}}{1-\frac{1}{x(z)\,x_{4,j}^{-}}}\comma\\
e^{i\chi_{2}(z)}=&\prod_{j=1}^{K_2}\frac{z-u_{2,j}-i}{z-u_{2,j}+i}\prod_{j=1}^{K_3}\frac{z-u_{3,j}+\frac{i}{2}}{z-u_{3,j}-\frac{i}{2}}\prod_{j=1}^{K_1}\frac{z-u_{1,j}+\frac{i}{2}}{z-u_{1,j}-\frac{i}{2}}\comma\\
e^{i\chi_{3}(z)}=&\prod_{j=1}^{K_2}\frac{z-u_{2,j}+\frac{i}{2}}{z-u_{2,j}-\frac{i}{2}}\prod_{j=1}^{K_4}\frac{x(z)-x_{4,j}^{+}}{x(z)-x_{4,j}^{-}}\comma\\
e^{i\chi_{4}(z)}=&\left[\frac{x^{+}(z)}{x^{-}(z)}\right]^{L}\prod_{j=1}^{K_4}\left(\frac{z-u_{4,j}-i}{z-u_{4,j}+i}\frac{1}{\sigma^{2}(x(z),x_{4,j})}\right)\\
&\quad \times \prod_{j=1}^{K_1}\frac{1-\frac{1}{x^{+}(z)\,x_{1,j}}}{1-\frac{1}{x^{-}(z)\, x_{1,j}}}\prod_{j=1}^{K_3}\frac{x^{+}(z)-x_{3,j}}{x^{-}(z)-x_{3,j}}\prod_{j=1}^{K_5}\frac{x^{+}(z)-x_{5,j}}{x^{-}(z)-x_{5,j}}\prod_{j=1}^{K_7}\frac{1-\frac{1}{x^{+}(z)\,x_{7,j}}}{1-\frac{1}{x^{-}(z)\,x_{7,j}}}
\end{aligned}
\eeq
Likewise the functions $\chi_{5}\,,\chi_6\,,$ and $\chi_7$ are defined by the simple replacement:
\beq
\chi_{8-a}(z) = \chi_{a}(z)\big{|}_{K_a\to K_{8-a},\,u_{a,j}\to u_{8-a,j}} 
\eeq
where $K_a$ is the occupation number at node $a$. In terms of these functions the Bethe equations can be simply stated as:
\beq
e^{i\chi_{a}(u_{a,k})}=(-1)^{a-1} \quad \text{with}\quad k=1,\cdots, K_a
\eeq
The sign on the right-hand side is $-1$ and $+1$ for bosonic and fermionic nodes respectively. 

\paragraph{The Gaudin norm and its factorization.}  The Gaudin matrix can be defined as a block matrix with $(ab)$ blocks given by: 
\beq
G_{ai,bj} \,=\, \frac{\partial\chi_{a}(u_{a,i})}{\partial u_{b,j}}%\bigg{|}_{z\to u_{a,i}}
\quad\text{with}\quad \substack{a,b\in\{1,\cdots,7\}\text{ and }\qquad\;\\i\in\{1,\cdots, K_a\},\;j\in\{1,\cdots, K_b\}}
\eeq
In practice the rank of the matrix is only given by the non-zero occupation numbers $K_a$. The Gaudin norm is computed as the determinant of this matrix: $\det G$. 

In our case of interest, the Bethe roots at each node come in pairs $\{u,-u\}$ and this results into a factorization of the Gaudin norm as:
\beq
\det\,G = \det\,G^{+}\times \det\,G^{-}
\eeq
The matrices $G^{\pm}$ also enter in the definition of the superdeterminant ${\rm SdetG}$ in \eqref{exact-1pt}. They are defined as:
\beq\label{eq:defGpm}
G_{ai,bj}^{\pm} \,=\,\frac{\partial\chi_{a}(u_{a,i})}{\partial u_{b,j}}\,\pm\, \frac{\partial\chi_{a}(u_{a,i})}{\partial \bar{u}_{b,j}}\,%\bigg{|}_{z\to u_{a,i}}
\quad\text{with }u_{b,j}\in \textbf{u}_b\text{ and }\bar{u}_{b,j}\equiv -u_{b,j}\in \bar{\textbf{u}}_b
\eeq
where the list of pairs of roots at node $b$ are split into two groups: $\textbf{u}_b \cup \bar{\textbf{u}}_b$. Due to this split, in general, the rank of the matrices $G^{\pm}$ is half the rank of the Gaudin matrix $G$. However we should also consider the special, but common, case of unpair roots: $u=0$ or $x(u)=0$\footnote{The case $x(u)=0$ happens only at fermionic nodes.}.  In this case, we modify the matrix $G^{+}$ by keeping the first term on the right-hand side of \eqref{eq:defGpm} for the elements where the zero root $u_{b,j}=0$ appears in the differential ``$\partial_{u_{b,j}}$". Furthermore, we modify the matrix $G^{-}$ 
 by omitting all the elements with $u_{a,i}=0$ or $u_{b,j}=0$ in \eqref{eq:defGpm}, resulting in a lower rank than that of $G^{+}$. In sections \ref{sec-examples}, \ref{sec:KonishiOneLoop} and \ref{sec:twist4oneloop} we present explicit examples of these matrices including unpaired roots at zero.

Finally, the superdeterminant in the (normalized) one-point function \eqref{exact-1pt} is defined by the ratio: 
\beq\label{eq:SdetG}
{\rm SdetG} = \frac{\det\,G^{+}}{\det\,G^{-}}\,.
\eeq
 
  \section{Tree-level condensates}\label{sec:treelevel}
    
 In this section and the next one we make contact between the formalism developed so far and perturbation theory, starting with tree level. To the first approximation,  the fields in an operator are just replaced by their classical expectation values projecting the operator's wavefunction onto a specific state. The overlap of this boundary state with the eigenstates of the dilatation operator can be computed using Bethe Ansatz  techniques \cite{deLeeuw:2015hxa}. The Coulomb branch, by this token, is associated to a particular state in the spin chain Hilbert space. We will construct this state explicitly and explore its properties in the $SO(6)$ sector consisting of scalar operators without derivatives.

The Lagrangian of the $\mathcal{N}=4$ super-Yang-Mills, written in the quasi-ten-dimensional form \cite{Brink:1976bc}, becomes
\begin{equation}\label{SYM-Lagrangian}
 \mathcal{L}=\frac{16\pi ^2N}{g^2}\mathop{\mathrm{tr}}
 \left\{
 -\frac{1}{2}\,F_{\mu \nu }^2+\left(D_\mu \Phi _i\right)^2
 +\frac{1}{2}\,\left[\Phi _i,\Phi _j\right]^2
 +i\bar{\Psi }\Gamma ^\mu D_\mu \Psi 
 +\bar{\Psi }\Gamma ^i\left[\Phi _i,\Psi \right]
 \right\},
\end{equation}
where $D_\mu \Phi =\partial _\mu \Phi -i\left[A_\mu ,\Phi \right]$, and the fermions are combined into  a single  ten-dimensional Majorana-Weyl spinor: $\Gamma ^{11}\Psi =\Psi $ and $\bar{\Psi }=\Psi ^\top C$. Here $\Gamma ^M $ are the 10d Dirac matrices and $C$ is the 10d charge-conjugation matrix: $C\Gamma ^M C^{-1} = -(\Gamma ^M)^\top$, $C^\top = -C$.  

To renormalize operators we use the background field method with the ghost and gauge-fixing terms:
\begin{equation}\label{gauge-fixing-BF}
 \mathcal{L}_{\rm gf+gh}=\frac{16\pi ^2N}{g^2}\mathop{\mathrm{tr}}\left\{
 -\left(\bar{D}_\mu A^\mu \right)^2
 +2\bar{D}_\mu \bar{c}D^\mu c
 \right\}.
\end{equation}
When the background field is switched off, these correspond to the usual Feynman gauge .

   \subsection{$SO(6)$ spin chain}
   
   The  $SO(6)$ sector consists of arbitrary combinations of the scalar fields under a single trace:
\begin{equation}\label{eq:wavefun}
 \mathcal{O}=\Psi ^{i_1\ldots i_L}\mathop{\mathrm{tr}}\Phi _{i_1}\ldots \Phi _{i_L}.
\end{equation}
The one-loop mixing of these operators is described by an integrable spin-chain Hamiltonian  \cite{Minahan:2002ve}:
\begin{equation}\label{so6-Ham}
 \Gamma =g^2\sum_{\ell=1}^{L}\left(2-2P_{\ell,\ell+1}+K_{\ell,\ell+1}\right),
\end{equation}
acting on the  wavefunction $\Psi $ by permuting and tracing nearest-neighbour indices ("sites of the spin chain"): $P_{ij}^{kl}=\delta _i^l\delta _j^k$, $K_{ij}^{kl}=\delta _{ij}\delta ^{kl}$. Beyond one loop the $SO(6)$ operators start mixing with fermions and then, at three loops, with gauge fields and derivatives. They only form a closed sector at one loop.

Contraction of indices defines a natural pairing of operators, which we denote by $\left\langle \Psi \right.\!\left|\Psi ' \right\rangle$ or $\left\langle \mathcal{O}\mathcal{O}'\right\rangle$. It is important to distinguish it from the true norm defined by the exact two-point function. At tree level the two are  simply related:
\begin{equation}\label{2pt-tree}
 \left\langle \mathcal{O}(x)\mathcal{O}'(0)\right\rangle_{\rm tree}\equiv 
 \frac{L(2g^2)^L}{x^{2L}}\,\left\langle \mathcal{O}\mathcal{O}'\right\rangle,
\end{equation}
where normalization is such that $\left\langle \Phi _i\Phi _j\right\rangle=\delta _{ij}$.
But at higher loops the two-point functions are not necessarly diagonal in the $SO(6)$ indices and are in general UV divergent. One can continue using the spin-chain norm provided the difference is appropriately accounted for in the physical correlation functions \cite{Ivanovskiy:2024vel}.  We also denote 
\begin{equation}
 \left\Vert\mathcal{O}\right\Vert^2\equiv \left\langle \mathcal{O}^\dagger \mathcal{O}\right\rangle.
\end{equation}

The one-point function at tree level amounts to evaluating the operator on the classical field (\ref{higgsvev}) and normalizing by the two-point function (\ref{2pt-tree}).  The result can be represented as a spin-chain overlap  \cite{Ivanovskiy:2024vel}:
\begin{equation}\label{tree-level<O>}
 \left\langle \mathcal{O}\right\rangle_{\rm tree}=\frac{\left\langle B\right.\!\left|\Psi  \right\rangle v^L}{2^{\frac{L}{2}}\sqrt{L}\,g^L\Vert  \Psi \Vert}\,
\end{equation}
with
\begin{equation}
 B_{i_{1}\ldots i_L}=n_{i_1}\ldots n_{i_L}.
\end{equation}
As already noted in \cite{Ivanovskiy:2024vel} the boundary state $\left\langle B\right|$ is integrable. The determinant formula  for its overlap with Bethe eigenstates is known \cite{deLeeuw:2019ebw} and can be regarded as a particular case of  the general construction of integrable overlaps 
\cite{Gombor:2023bez,Gombor:2024iix,gombor2025der}. 

The Bethe equations for the spectrum of the Hamiltonian (\ref{so6-Ham})  are 
\begin{equation}\label{1loop-BAE}
 \,{\rm e}\,^{i\chi _{aj}}\equiv \left(\frac{u_{aj}-\frac{iq_a}{2}}{u_{aj}+\frac{iq_a}{2}}\right)^L\prod_{bk}^{}\frac{u_{aj}-u_{bk}+\frac{iM_{ab}}{2}}{u_{aj}-u_{bk}-\frac{iM_{ab}}{2}}=-1,
\end{equation}
where $M_{ab}$ is the Cartan matrix of $\mathfrak{so}(6)$ and $q_a$ are the Dynkin labels of the defining representation:
\begin{equation}
 M=\begin{bmatrix}
 2  & -1  & 0 \\ 
  -1 & 2 &  -1 \\ 
  0 & -1 & 2 \\ 
 \end{bmatrix},
 \qquad 
 q=\begin{bmatrix}
 0  \\ 
  1 \\ 
  0 \\ 
 \end{bmatrix},
 \qquad 
 \stackrel{}{\ocircle}\!\!\!-\!\!\! -\!\!\!-\!\!\!\stackrel{1}{\ocircle}\!\!\!-\!\!\! -\!\!\!-\!\!\!\stackrel{}{\ocircle}.
\end{equation}
The one-loop anomalous dimension is equal to
\begin{equation}
 \gamma =2g^2\sum_{aj}^{}\frac{q_a}{u_{aj}^2+\frac{q_a^2}{4}}\,.
\end{equation}

The states with non-zero overlap $\left\langle B\right.\!\left|\mathbf{u} \right\rangle$ must be invariant under $u_{aj}\rightarrow -u_{aj}$ with the Bethe roots forming symmetric pairs $\left\{u_{aj},-u_{aj}\right\}$ or containing unpaired roots at zero. For those  states,
\begin{equation}\label{SO(6)-overlap}
 \left\langle \mathcal{O}_{\mathbf{u}}\right\rangle_{\rm tree}
 =\left(\frac{v}{2g}\right)^L\sqrt{\frac{1}{L}\,\,
 \frac{Q_4\left(0\right)Q_4\left(\frac{i}{2}\right)}{Q_3\left(0\right)Q_3\left(\frac{i}{2}\right)Q_5\left(0\right)Q_5\left(\frac{i}{2}\right)}\,\,\frac{\det G^+}{\det G^-}
 }\,.
\end{equation}
To facilitate comparison with  $PSU(2,2|4)$, to be discussed shortly, the nodes of the  Dynkin diagram are labelled $3,4,5$. The Q-functions are defined with zero roots excluded:
\begin{equation}
 Q_a(u)=\prod_{j=1}^{N_a/2}\left(u^2_{aj}-u^2\right).
\end{equation}
Eventual zeros at nodes $a_{\alpha}$ contribute to $G^+$ but not $G^-$:
\begin{eqnarray}\label{1-loop-Gaudin}
 G^\pm_{aj,bk}&=&\left(\frac{Lq_a}{u_{aj}^2+\frac{q_a^2}{4}}-\sum_{cl}^{}K^+_{aj,cl}
 -\frac{1}{2}\sum_{\alpha }^{}\,K^+_{aj,a_\alpha 0}\right)\delta _{ab}\delta _{jk}+K^\pm_{aj,bk} \, ,
\nonumber \\
 G^+_{aj,\alpha }&=&\frac{1}{\sqrt{2}}\,K^+_{aj,a_\alpha 0}=G^+_{\alpha ,aj} \, ,
\nonumber \\
G^+_{\alpha \beta }&=&\left(\frac{4L}{q_{a_\alpha }}-\sum_{cl}K^+_{a_\alpha 0,cl}-\sum_{\gamma }^{}\frac{4}{M_{a_\alpha a_\gamma }}\right)
 \delta _{\alpha \beta }+\frac{4}{M_{a_\alpha b_\beta }}\,.
\end{eqnarray}
Here 
\begin{equation}
 K_{aj,bk}^\pm=\frac{M_{ab}}{(u_{aj}-u_{bk})^2+\frac{M_{ab}^2}{4}}\pm
 \frac{M_{ab}}{(u_{aj}+u_{bk})^2+\frac{M_{ab}^2}{4}}\,.
\end{equation}
The entries of the Gaudin matrices are formally divergent for $q_a=0$ or $M_{ab}=0$. Whenever this happens the divergent expressions  $1/q_a$ and $1/M_{ab}$ should be  set to zero by default. 

\begin{figure}[t]
\centering
 \centerline{\includegraphics[width=3.2cm]{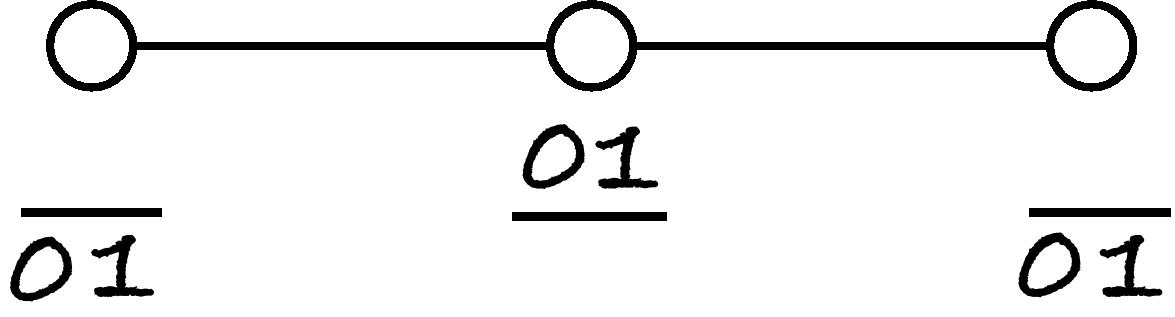}}
\caption{\label{Dynkin-SO6}\small  A  graphic representation of the $SO(6)$ overlap~ (\ref{SO(6)-overlap}).}
\end{figure}

The determinant formula can be represented pictorially following the graphic conventions introduced in \cite{Kristjansen:2020vbe}: the ratio of Q-functions $\frac{Q_a(is_1/2)\ldots Q_a(is_n/2)}{Q_a(ir_1/2)\ldots Q_a(ir_m/2)}$, that appears as a prefactor in the overlap formula, is assigned a label $\frac{s_1\ldots s_n}{r_1\ldots r_m}$  at  the $a$th node of the Dynkin diagram. The analytic expression (\ref{SO(6)-overlap}) then corresponds to fig.~\ref{Dynkin-SO6}.

 The $SO(6)$ overlap formula should be equivalent to the non-perturbative expression from the previous section  when the latter is restricted to one loop and to the $SO(6)$ subsector. The $SO(6)$ reduction is a  non-trivial step because  the non-perturbative Bethe equations do not fit well with $SO(6)$. A faithful comparison requires fermionic dualities to shuffle between different gradings of the $\mathfrak{psu}(2,2|4)$ superalgebra \cite{Kristjansen:2020vbe,Kristjansen:2021xno}. Before going into details we pause for illustrating the formalism on a number of examples.

\subsection{Examples}\label{sec-examples}

\paragraph{Konishi} The Konishi operator
\begin{equation}
 K=\mathop{\mathrm{tr}}\Phi _i\Phi _i
\end{equation}
corresponds to the configuration of Bethe roots 
\begin{equation}\label{Konishi-Bethe-roots}
 \stackrel{\left\{0\right\}}{\ocircle}\!\!\!\!-\!\!\! -\!\!\!-\!\!\!\!\!\!\!\stackrel{\left\{u,-u\right\}}{\ocircle}\!\!\!\!\!\!\!-\!\!\! -\!\!\!-\!\!\!\!\stackrel{\left\{0\right\}}{\ocircle}
\end{equation}
with $u=1/\sqrt{12}$, yielding $\gamma _K=12g^2$. The Gaudin determinants (\ref{1-loop-Gaudin}) evaluated on these data give
\begin{equation}
 \det G^+=
 \det\begin{bmatrix}
\frac{2}{u^2+\frac{1}{4}}   &  \frac{\sqrt{2}}{u^2+\frac{1}{4}}   & 0 \\ 
  \frac{\sqrt{2}}{u^2+\frac{1}{4}}   & \frac{L+2}{u^2+\frac{1}{4}}  &   \frac{\sqrt{2}}{u^2+\frac{1}{4}}  \\ 
 0   &  \frac{\sqrt{2}}{u^2+\frac{1}{4}}  & \frac{2}{u^2+\frac{1}{4}}  \\ 
 \end{bmatrix}
 =\frac{4L}{\left(u^2+\frac{1}{4}\right)^3}\,,
 \qquad 
 \det G^-=\frac{L+1}{u^2+\frac{1}{4}}\,,
\end{equation}
and the overlap formula boils down to
\begin{equation}\label{Konshi-like-1pt}
 \left\langle K\right\rangle=\left(\frac{v}{2g}\right)^L\sqrt{\frac{4u^2}{(L+1)\left(u^2+\frac{1}{4}\right)}}=\frac{v^2}{4\sqrt{3}\,g^2}\,.
\end{equation}
Comparing to the direct calculation we notice that the tree-level  vev  of the  operator is just $v^2$, while the norm is dictated by the two-point function $2\cdot 6\cdot (2g^2/x^2)^2$ and indeed accounts for the normalization constant as it appears in (\ref{Konshi-like-1pt}).

\paragraph{Dimension-3 isovector} The Konishi scalar is a rare example of unprotected operator with no operator mixing. Another state that shares this property is the isotopic vector of bare dimension three: $\mathop{\mathrm{tr}}\Phi _i\Phi _i\Phi _j$. We consider a particular component, the highest weight of the vector multiplet:
\begin{equation}
 V=\mathop{\mathrm{tr}}\Phi _i\Phi _iZ.
\end{equation}
This operator  corresponds to 
the solution of the Bethe equations with two momentum-carrying roots and two zeros on the auxiliary nodes, just like for Konishi, but now with $u=1/2$, giving  $\gamma _V=8g^2$ for the anomalous dimension. The one-point function also follow from (\ref{Konshi-like-1pt}):
\begin{equation}\label{isovector-tree}
 \left\langle V\right\rangle=\frac{1}{\sqrt{2}}\left(\frac{v}{2g}\right)^3,
\end{equation}
and again agrees with the direct calculation, because the norm now is $(6\cdot 2+2\cdot 2)\left(2g^2/x^2\right)^3$.

\paragraph{Two-magnon overlaps} A natural generalization of the Konishi operator and the dimension-3 scalar  to any length is the $SO(4)$-singlet BMN operator \cite{Berenstein:2002jq}. The exact form of the BMN operator at one loop is \cite{Beisert:2002tn}
\begin{equation}\label{Konishi-nL}
 \mathcal{O}_{nL}
 =-\sum_{l=0}^{L-2}\cos\frac{(2l+3)\pi n}{L+1}\,\mathop{\mathrm{tr}}\Phi _iZ^l\Phi _iZ^{L-l-2},\qquad n=1,\ldots, [L/2].
\end{equation}
In particular,  $K=\mathcal{O}_{12}$ and $V=\mathcal{O}_{13}$.
The Bethe roots of the BMN singlets have the same structure as (\ref{Konishi-Bethe-roots})  with   \cite{Minahan:2002ve}
\begin{equation}\label{u-BMN}
 u=\frac{1}{2}\,\cot\frac{\pi n}{L+1}\,,
\end{equation}
which gives for the anomalous dimension:
\begin{equation}
 \gamma _{nL}=16g^2\sin^2\frac{\pi n}{L+1}\,.
\end{equation}
The condensates of the BMN singlets on the Coulomb branch were calculated in \cite{Ivanovskiy:2024vel}:
\begin{equation}
 \left\langle \mathcal{O}_{nL}\right\rangle=\frac{2\cos\frac{\pi n}{L+1}}{\sqrt{L+1}}\,\left(\frac{v}{2g}\right)^L.
\end{equation}
It is easy to see  that the determinant formula gives the same result once $u$ from (\ref{u-BMN}) is substituted into (\ref{Konshi-like-1pt}). 

\paragraph{Dimension-four singlets} States with four magnons appear first at length four, where they describe mixing of two independent $SO(6)$ singlets. The linear combinations that diagonalize the dilatation operator are
\begin{equation}\label{Opm}
 \mathcal{O}_\pm=\mathop{\mathrm{tr}}
 \left(
 \Phi _i\Phi _i\Phi _j\Phi _j+\frac{5\mp\sqrt{41}}{4}\,\Phi _i\Phi _j\Phi _i\Phi _j
 \right).
\end{equation}
These operators are eigenstates of the mixing matrix (\ref{so6-Ham}) with the eigenvalues $\gamma _\pm=(13\pm\sqrt{41})g^2$.
They have appeared in many studies \cite{Beisert:2003jj,Beisert:2003yb,Grossardt:2010xq,Georgiou:2012zj,Caron-Huot:2022sdy,Eden:2023ygu}, and in particular their Coulomb-branch one-point functions are known from \cite{Ivanovskiy:2024vel}:
\begin{equation}\label{<O+->}
 \left\langle \mathcal{O}_\pm\right\rangle=\frac{1}{2}\left(\frac{v}{2g}\right)^4
 \sqrt{\frac{1}{2}\mp\frac{7}{6\sqrt{41}}}\,.
\end{equation}
The solution of the Bethe equations for $\mathcal{O}_\pm$ contains four momentum-carrying and four auxiliary roots:
\begin{equation}\label{rootsOpm}
 \stackrel{\left\{w,-w\right\}}{\ocircle}\!\!\!\!\!\!\!\!-\!\!\!-\!\!\!-\!\!\!-\!\!\!-\!\!\!-\!\!\!-\!\!-\!\!\!\!\!\!\!\!\!\!\!\!\stackrel{\left\{u,-u,v,-v\right\}}{\ocircle}\!\!\!\!\!\!\!\!\!\!\!\!-\!\!-\!\!\!-\!\!\!-\!\!\!-\!\!\!-\!\!\!-\!\!\!-\!\!\!\!\!\!\!\!\stackrel{\left\{w,-w\right\}}{\ocircle}
\end{equation}
where
\begin{eqnarray}\label{LO-B-roots}
 u^2&=&\frac{3\pm2\sqrt{41}+\sqrt{\pm 42\sqrt{41}-82}}{60}
\nonumber \\
 v^2&=&\frac{3\pm2\sqrt{41}-\sqrt{\pm 42\sqrt{41}-82}}{60}
\nonumber \\
w^2&=&\frac{\pm\sqrt{41}-1}{20}\,.
\end{eqnarray}
While the Gaudin factors $G^\pm$ are fairly complicated $4\times 4$ matrices, their ratio simplifies:
\begin{equation}
 \frac{\det G^+}{\det G^-}
 =\frac{1}{10}\left(9\pm\frac{51}{\sqrt{41}}\right),
\end{equation}
and, when multiplied by the appropriate Q-functions  and substituted into (\ref{SO(6)-overlap}), reproduces (\ref{<O+->}). 

\paragraph{Higher-dimension $SO(6)$ operators} We have also tested the tree-level formula \eqref{SO(6)-overlap} for higher-dimension operators. By making a direct computation of the overlap, using the explicit wavefunctions $\Psi$ defined in \eqref{eq:wavefun}. These eigenstates can be obtained by a direct diagonalization of the one-loop dilatation operator in \eqref{so6-Ham} or by using the 
$SO(6)$ coordinate Bethe Ansatz given in appendix E of
\cite{Basso:2017khq}. We obtained a perfect match for operators with $SO(6)$ charges as high as $L=8$ with $\bar{\textbf{K}}_{SO(6)} = \{4,8,4\}$.

\subsection{$PSU(2,2|4)$ and fermionic dualties}
\label{subsec:duality}
\begin{figure}[t]
\centering
 \centerline{\includegraphics[width=9cm]{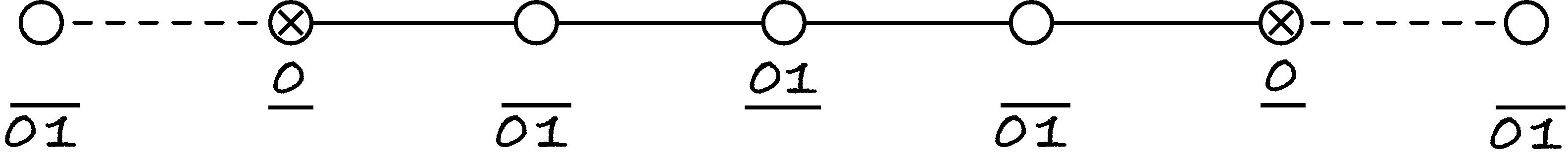}}
\caption{\label{Dynkin-BeautyG}\small The $PSU(2,2|4)$ overlap in the Beauty grading.}
\end{figure}

\begin{figure}[t]
\centering
 \centerline{\includegraphics[width=9cm]{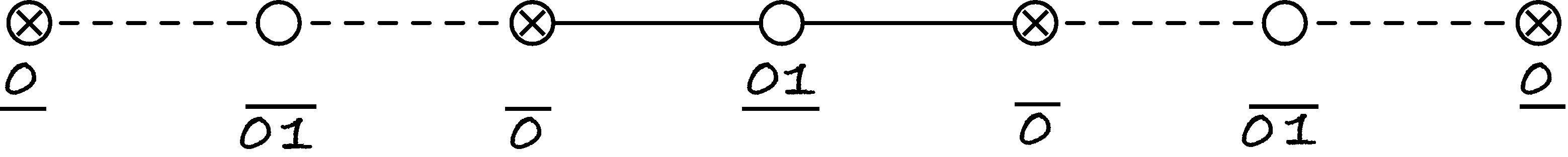}}
\caption{\label{Dynkin-AltG}\small The overlap in the alternate grading, as obtained from the weak-coupling limit of the non-perturbative one-point function. The label $\frac{s_1\ldots s_n}{r_1\ldots r_m}$  below the $a$th node stands for the factor  $\frac{Q_a(is_1/2)\ldots Q_a(is_n/2)}{Q_a(ir_1/2)\ldots Q_a(ir_m/2)}$ in the overlap formula \eqref{tree-overlap}.
}
\end{figure}

The $PSU(2,2|4)$ extension of the $SO(6)$ overlap (\ref{SO(6)-overlap}) is shown in fig.~\ref{Dynkin-BeautyG}. For comparison, the weak-coupling limit of the exact one-point function (\ref{exact-1pt2})  is
\begin{equation}\label{tree-overlap}
 \left\langle \mathcal{O}\right\rangle_v
 =\mathbbm{C}_{\mathbf{K}}\left(\frac{v}{2g }\right)^L\sqrt{\frac{1}{L}\,\,
 \frac{Q_1(0)Q_4\left(0\right)Q_4\left(\frac{i}{2}\right)Q_7(0)}{Q_2(0)Q_2(\frac{i}{2})Q_3\left(0\right)Q_5\left(0\right)Q_6(0)Q_6\left(\frac{i}{2}\right)}\,
 \,\frac{\det G^+}{\det G^-}
 }\,.
\end{equation}
To arrive at this formula we replaced $x_{aj} \rightarrow u_{aj}/g$ for all roots, see (\ref{zhuk}), and specialized to the case of $K_{1}=K_{3}=K_{5}=K_{7}$ that are relevant for the analysis below.
Fig.~\ref{Dynkin-AltG} is a pictorial representation of the overlap that follows from the exact one-point function. The two formulas cannot be compared immediately because they correspond to different fermionic gradings of the $PU(2,2|4)$ Dynkin diagram.

The equivalence between fig.~\ref{Dynkin-BeautyG} and fig.~\ref{Dynkin-AltG} follows from fermionic dualities that transform the Bethe Ansatz equations from one grading to another \cite{Tsuboi:1998ne,Beisert:2005di}. Fermionic duality is part of a larger web of QQ-relations \cite{Kazakov:2007fy} and as such plays an important role in the Quantum Spectral Curve formalism \cite{Gromov:2014caa}. Importantly for us, the Gaudin superdeterminant transforms covariantly under the duality, with a simple combination of the Q-functions as a Jacobian  \cite{Kristjansen:2020vbe}. Because of that the overlap formula preserves its general form  in the new frame under some favorable conditions. If those are not met, the overlap will not have any simple representation in the new grading.
However, if at least some of  the gradings are allowed, consistency with fermionic duality imposes strong constraints and alone can fix the supersymmetric extension of the known bosonic overlap.

\begin{figure}[t]
\centering
 \centerline{\includegraphics[width=8cm]{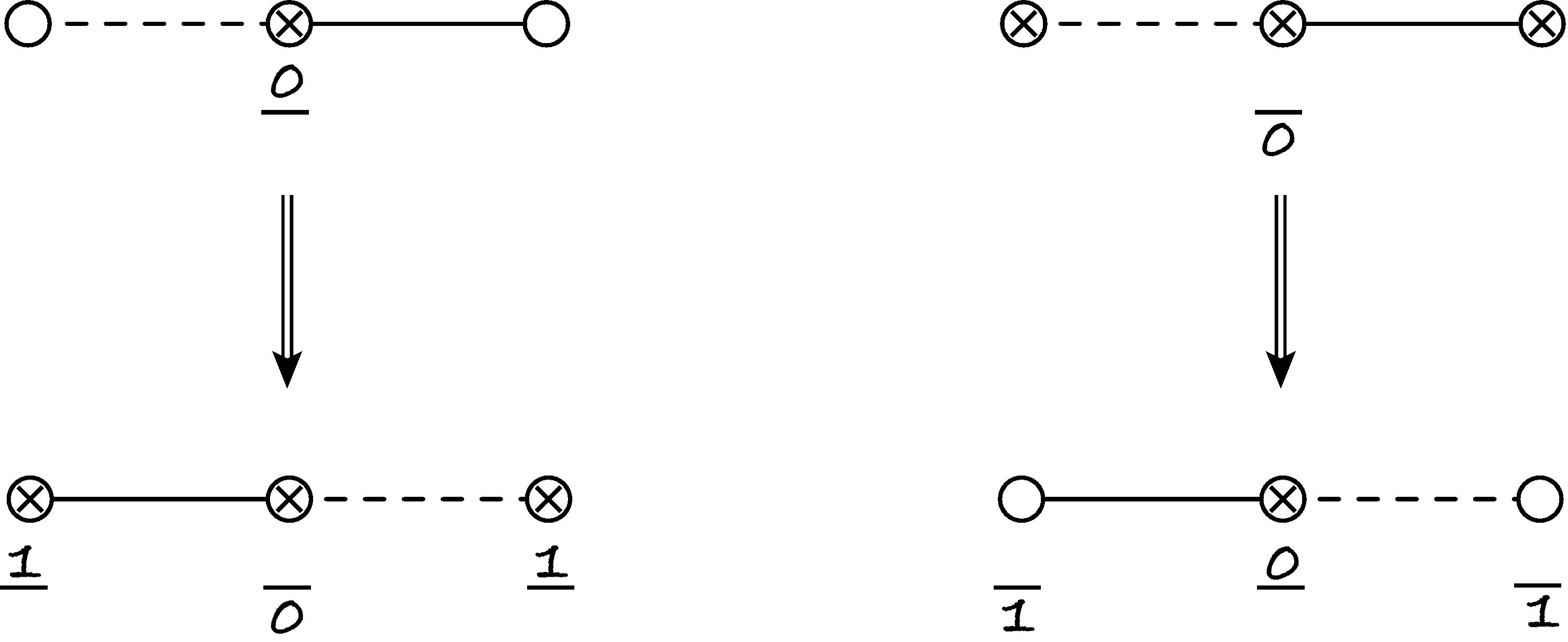}}
\caption{\label{D-Rules}\small Fermionic duality transformations.}
\end{figure}

An elementary move of the duality is a change of variables on a fermion node of the Dynkin diagram that effectively flips grading of its nearest neighbours \cite{Tsuboi:1998ne}. 
As for the overlap formulas, there are two possibilities:  the only Q-function allowed on the active node is either  $Q_a(0)$ or $Q_a(0)^{-1}$, which gets reversed in the dual frame: $Q_a(0)$ becomes $Q_a(0)^{-1}$ and $Q_a(0)^{-1}$ becomes $Q_a(0)$, with an extra factor of $Q_{a+ 1}(i/2)Q_{a- 1}(i/2)$ appearing in the numerator or denominator \cite{Kristjansen:2020vbe}. These rules are  illustrated in fig.~\ref{D-Rules}.
If the dependence on $Q_a(u)$ is more complicated than $Q_a(0)^{\pm 1}$,  duality generates a mess.

\begin{figure}[t]
\centering
 \centerline{\includegraphics[width=9cm]{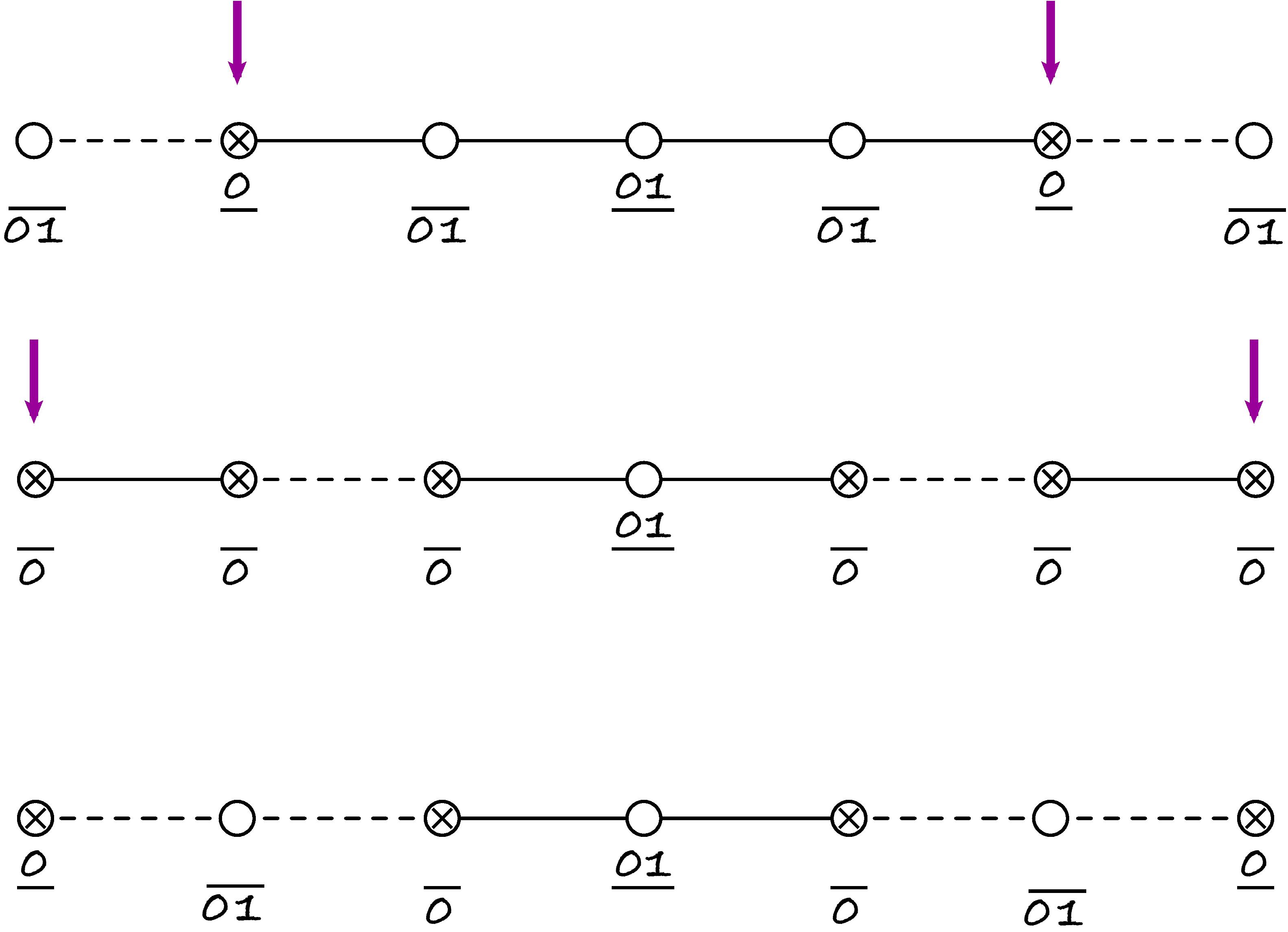}}
\caption{\label{Dualize}\small Dualizing from Beauty to alternate grading. The active nodes are marked by arrows.}
\end{figure}

The chain of dualities linking the Beauty diagram with the alternating one is shown in fig.~\ref{Dualize}. We immediately notice that  the duality would crash on the first step without  $0/$ on the 2nd node and without $/01$ on the first node the second step would fail, hence fermionic dualities uniquely fix the supersymmetric extention of the $SO(6)$ overlap\footnote{Though $/0$ is also formally allowed, it does not cancel $/1$ on the third node impeding duality transformations there. We have not dualized the 3rd node, but with the current choice the duality chain can be continued further in that direction.}. Another consistency check is the symmetric depdence on $u_1$ and $1/u_3$. In the asymptotic Bethe ansatz the 1st and 3rd nodes are combined into one, and the symmetry between the respective Q-functions is imperative for an all-loop extension.

The overlap formulas apply as written only to primary operators, but
states that are primary in one grading in general become descendants in the other. This is the case for all operators from sec.~\ref{sec-examples}, they are primaries of $SO(6)$ and hence of $PSU(2,2|4)$ in the Beauty grading, but are descendants in the alternate grading. Only the former can be faithfully generalized to higher loops and to compute the one-point functions of operators at hand beyond the leading order we need an overlap formula for descendants. 

The one-point functions of descendants are determined by symmetries, and in principle can be fixed by the Wigner-Eckart theorem. The action of symmetry generators produces an overall  Clebsch–Gordan-like prefactor that depends on the quantum numbers of the state but not on the Bethe roots\footnote{Beyond tree level this is no longer true, see \cite{Gombor:2024api} and the next subsection. The full-fledged symmetry pre-factors have been worked out only for $SU(2)$ \cite{deLeeuw:2017dkd},  for $PSU(2,2|4)$ they are known in one particular example \cite{Gombor:2024api}.}. We will not follow this strategy here but will instead reconstruct the symmetry prefactors from the fermionic duality.

The duality transformation applied at the $a$th node changes its occupation number:
\begin{equation}
 K_a\rightarrow \widetilde{K}_a= K_{a-1}+K_{a+1}-K_a-1.
\end{equation}
The Bethe roots on the other nodes  remain intact. Apart from  the Q-functions, the Jacobian produces a combinatorial factor  \cite{Kristjansen:2020vbe}, which is different for the two possible moves in fig.~\ref{D-Rules}:
\begin{align}
 {\tt move~1}:&\qquad (K_{a-1}-K_{a+1})^{-\frac{1}{2}}
\nonumber \\
 {\tt move~2}:&\qquad  (K_{a-1}-K_{a+1})^{\frac{1}{2}}.
\nonumber 
\end{align}
Knowing the symmetry factor in one grading we can trace these combinatorial factors along the duality chain  and thus reconstruct the prefactor in a different grading. 

We work out this procedure for the $SO(6)$ operators, which are primaries of $PSU(2,2|4)$ in the Beauty grading. In that grading the $SO(6)$ sector is characterized by the occupation numbers 
$$(0,0,K_3,K_4,K_5,0,0).$$ Tracing how the dualities in fig.~\ref{Dualize} act we find the occupation numbers  in the alternate grading: $$(K_2-1,K_2,K_2+1,K_4,K_6+1,K_6,K_6-1),$$
where for later convenience we relabelled $K_3\rightarrow K_2+1$, $K_5\rightarrow K_6+1$. 
Collecting the combinatorial  factors at each step we also find the overall symmetry pre-factor:
\begin{equation}\label{so6-tree-symmfact}
 \mathbbm{C}_{\mathbf{K}}=\sqrt{\frac{K_2K_6}{(K_2+1)(K_6+1)}}\,.
\end{equation}
Taken literally, this result implies vanishing overlap for $K_2=0$ or $K_6=0$, but this makes no sense. The second step of the duality acts trivially for states with no $K_2$ or $K_6$ roots (the empty outermost node remains empty), the Bethe roots remain the same and the Gaudin factors simply do not change, so the vanishing numerator should just be replaced by one. The symmetry factor for Konishi-like operators (\ref{Konishi-Bethe-roots}) is thus
\begin{equation}
\label{degenerate-tree-symmfact}
 \mathbbm{C}_{\mathbf{K}}=\frac{1}{\sqrt{(K_2+1)(K_6+1)}}=1\qquad ({\rm for~}K_2=0=K_6)\,.
\end{equation}

To conclude, the  one-point function of a descendant operator will have a combinatorial pre-factor:
\begin{equation}\label{eq:su2treelevel}
 \left\langle \mathcal{O}\right\rangle
 =\mathbbm{C}_{\mathbf{K}}\left(\frac{v}{2g }\right)^L\sqrt{\frac{1}{L}\,\,
 \frac{Q_1(0)Q_4\left(0\right)Q_4\left(\frac{i}{2}\right)Q_7(0)}{Q_2(0)Q_2(\frac{i}{2})Q_3\left(0\right)Q_5\left(0\right)Q_6(0)Q_6\left(\frac{i}{2}\right)}\,
 \,\frac{\det G^+}{\det G^-}
 }\,.
\end{equation}
For the $SO(6)$ primaries the symmetry factor is given by (\ref{so6-tree-symmfact}), for other states it can be worked out in a similar fashion or extracted from the Wigner-Eckart theorem.

A good illustration of a difference between primaries and descendants is the dimension-four singlet (\ref{Opm}). Its roots in the Beauty grading are given by (\ref{rootsOpm}), (\ref{LO-B-roots}), while in the alternate grading two additional zeros on the nodes 2 and 6 should be added: \begin{equation}\label{alter-rootsOpm}
 \stackrel{}{\otimes}\!-\!\!\!\!\!\!\!\!-\!\!\!-\!\!\!-\!\!\!-\!\!\!-\!\!\!-\!\!\!-\!\!-\!\!\!\!\!\!\!\!-
 \stackrel{\left\{0\right\}}{\ocircle}\!-\!\!\!\!\!\!\!\!-\!\!\!-\!\!\!-\!\!\!-\!\!\!-\!\!\!-\!\!\!-\!\!-\!\!\!\!\!\!\!\!\stackrel{\left\{w,-w\right\}}{\otimes}\!\!\!\!\!\!\!\!-\!\!\!-\!\!\!-\!\!\!-\!\!\!-\!\!\!-\!\!\!-\!\!-\!\!\!\!\!\!\!\!\!\!\!\!\stackrel{\left\{u,-u,v,-v\right\}}{\ocircle}\!\!\!\!\!\!\!\!\!\!\!\!-\!\!-\!\!\!-\!\!\!-\!\!\!-\!\!\!-\!\!\!-\!\!\!-\!\!\!\!\!\!\!\stackrel{\left\{w,-w\right\}}{\otimes}-\!\!\!\!\!\!\!\!\!\!\!\!\!\!-\!\!\!-\!\!\!-\!\!\!-\!\!\!-\!\!\!-\!\!\!-\!\!\!-\!\!\!\stackrel{\left\{0\right\}}{\ocircle}--\!\!\!\!\!\!\!\!\!\!\!\!\!\!\!-\!\!\!-\!\!\!-\!\!\!-\!\!\!-\!\!\!-\!\!\!-\!\!\!-\!\!\!\stackrel{}{\otimes}
\end{equation}
The overlap formula (\ref{tree-overlap}) with these data gives the result twice as big as  (\ref{<O+->}), and the symmetry factor $\sqrt{1\cdot 1/2\cdot 2}$ is necessary to correct for  this discrepancy.

At higher loops the symmetry  pre-factor starts to depend on the Bethe roots \cite{Gombor:2024api}, and we do not know any systematic procedure to compute it, nor do we try to derive it from the first principles. Instead we outline a conjecture on how to extend the tree-level pre-factor  to higher orders in perturbation theory. We will extensively check this conjecture by explicit one-loop computations for the $SO(6)$ sector.

\subsection{Descendants}
\label{subsec:descendant}
In the Bethe-Ansatz language the descendants are associated with roots at infinity that carry zero spin-chain momentum. Indeed, adding a zero-momentum magnon  is equivalent to acting on the state by a symmetry generator. An obvious idea then is to augment the state with an appropriate number of extra roots (we call them virtual roots), and then  take the limit $u_{aj}^{\rm virt.}\rightarrow \infty $. By inspecting the descendant overlaps in the Heisenberg model, where they are known explicitly, we find that this idea does not quite work, not in the most direct way at least  (see appendix~\ref{OD-Heisenberg}). We observe nonetheless that the true descendant overlaps and primary overlaps with virtual roots  depend on the same type of data. Namely, they depend only on two numbers, the number of virtual roots $n_a$ and the number of vacancies 
\begin{equation}\label{nu-def}
 \nu _a=\lim_{v\rightarrow \infty }v^2\,\frac{\partial \chi _a(v)}{\partial v}\,,
\end{equation}
where $\chi _a$ is the phase of the Bethe equation for the roots of type $a$. This last parameter defines the maximal  number of virtual roots a given state can accommodate. Adding a root at infinity to the Gaudin determinant results a combinatorial factor that depends precisely on the same parameters, and we conjecture that the true overlap also depends only on the number of virtual roots $n_a$ and the number of vacancies $\nu _a$.

At one loop, $\nu _a$ are integers depending on the filling fractions:
\begin{equation}
 \nu _a=Lq_a-\sum_{b}^{}M_{ab}K_b.
\end{equation}
In light of the preceding discussion we re-interpret $K_2$ and $K_6$ in (\ref{so6-tree-symmfact}), (\ref{degenerate-tree-symmfact}) as correspondingly $\nu _1$ and $\nu _7$. The difference between (\ref{so6-tree-symmfact}) and (\ref{degenerate-tree-symmfact}) can be attributed to the structure of virtual roots. For a generic state in the $SO(6)$ sector
\begin{equation}
 n_a=(2,1,0,0,0,1,2),
\end{equation}
but for the two-magnon states (\ref{Konishi-Bethe-roots}) the number of virtual roots is smaller (the purely scalar descendant is obtained by levelling the number of roots on the outer nodes with $K_3$ and $K_5$):
\begin{equation}
 n_a=(1,1,0,0,0,1,1).
\end{equation}

In the all-loop Bethe equations, the number of vacancies as defined in (\ref{nu-def}) becomes dynamical:
\begin{equation}\label{dyn-K}
 -\nu _1
 =K_2+ig\sum_{j=1}^{K_4}\left(\frac{1}{x_{4j}^+}-\frac{1}{x_{4j}^-}\right)\equiv \mathbbm{K}_2,
 \qquad 
 -\nu _7
 =K_6+ig\sum_{j=1}^{K_4}\left(\frac{1}{x_{4j}^+}-\frac{1}{x_{4j}^-}\right)\equiv \mathbbm{K}_6,
\end{equation}
and we conjecture that the all-loop symmetry factors are obtained by simply replacing $K_2$ and $K_6$ with their all-loop counterparts:
\begin{align}
  \mathbbm{C}_{\mathbf{K}}&=\sqrt{\frac{\mathbbm{K}_2\mathbbm{K}_6}{(\mathbbm{K}_2+1)(\mathbbm{K}_6+1)}}\qquad ({\rm for~}n_1=2=n_7) \label{eq:CKn2}
\\
 \mathbbm{C}_{\mathbf{K}}&=\frac{1}{\sqrt{(\mathbbm{K}_2+1)(\mathbbm{K}_6+1)}}\qquad ({\rm for~}n_1=1=n_7). \label{eq:CKn1}
\end{align}
By inspecting (\ref{dyn-K}) we see that the occupation numbers get shifted by half the anomalous dimension:
\begin{equation}
 \mathbbm{K}_{2,6}=K_{2,6}+\frac{\gamma }{2}\,.
\end{equation}
Quantum deformation of this type first appeared in the three-point functions derived from the separation of variables \cite{Bercini:2022jxo}. The single known  $PSU(2,2|4)$ symmetry factor  \cite{Gombor:2024api} also contains integer labels shifted by the anomalous dimension. We take this as an indirect evidence for our conjecture. 

 \section{One-loop corrections}\label{sec:1loop}

In this section we compare the one-loop approximation of the asymptotic formula with  independent field-theory computations wherever the latter are available. We start by summarizing the  field-theory results.

\subsection{Experimental data}

The general one-loop formula for scalar operators was derived in \cite{Ivanovskiy:2024vel} and  looks deceptively simple:
\begin{equation}\label{1-loop-gen}
 \left\langle \mathcal{O}\right\rangle_{\rm 1-loop}
 =\frac{1}{2^{\frac{L}{2}}\sqrt{L}\,g^L}\frac{\left\langle \mathcal{O}\right\rangle_{\rm tree}}{\left\langle \mathcal{O} \mathcal{O}\right\rangle_{\rm tree}^{\frac{1}{2}}}
 \left[
 1+\left(\gamma _E+\ln\frac{v}{2}\right)\gamma 
 \right].
\end{equation}
The expectation value and the norm on the right-hand side are evaluated at tree level, with the normalization conventions (\ref{2pt-tree}), just like in  the tree-level formula. It might seem that the only effect of one-loop corrections is a simple overall prefactor, where the  anomalous dimension  $\gamma $ is actually mandated by the $v^\Delta $ scaling of the condensate.
This is not quite true. The most complicated contribution comes not from the loop diagrams as such but from the operator mixing implicit in the formula. The requisite accuracy is two loops, at two loops the $SO(6)$ sector is no longer closed and  scalar operators start mixing with fermions, and to faithfully compute the operator norm we need to know the  form of  the operator at the two-loop accuracy including its fermion components. A general recipy to find the fermion component does not exist, but in a number  of examples the two-loop exact eigenstates have been worked out \cite{Georgiou:2012zj} by combining tour de  force diagrammatic calculations \cite{Georgiou:2011xj} with supersymmetry \cite{Georgiou:2008vk,Georgiou:2009tp}. Their one-point functions can then be evaluated explicitly \cite{Ivanovskiy:2024vel}. To make presentation self-contained we list these known results below.

\paragraph{Konishi and dimension-3 isovector.} These are the only two operators that do not mix with anything and their one-point functions are obtained by multipling the tree-level result (\ref{Konshi-like-1pt}) or (\ref{isovector-tree}) with the one-loop prefactor from (\ref{1-loop-gen}):
\begin{align}
 \left\langle K\right\rangle_{\rm 1-loop}&=\frac{v^2}{4\sqrt{3}\,\lambda }
 \left[1+12g^2\left(\gamma_E +\ln\frac{v}{2}\right)\right]
 \\
  \left\langle V\right\rangle_{\rm 1-loop}&=\frac{v^3}{8\sqrt{2}\,g^3}
 \left[1+8g^2\left(\gamma_E +\ln\frac{v}{2}\right)\right].
\end{align}
In all other cases we need to resolve the two-loop mixing first.
 
 \paragraph{Two-magnon operators.} The two-loop mixing of the BMN singlets (\ref{Konishi-nL})  was worked out in \cite{Georgiou:2012zj} and  their one-point functions were found to be  \cite{Ivanovskiy:2024vel}   
 \begin{equation}\label{av-int}
 \left\langle \mathcal{O}_{nL}\right\rangle_{\rm 1-loop}=\frac{2\cos\frac{\pi n}{L+1}}{\sqrt{L+1}}\,\left(\frac{v}{2g}\right)^L
 \left(
 1+16g^2C_{nL}\sin^2\frac{\pi n}{L+1}
 \right).
 \end{equation}
with
\begin{equation}\label{CnL}
 C_{nL}=\gamma_E +\ln\frac{v}{2}+\frac{1}{4}\,\,\frac{L-2}{L+1}\,
 \cos\frac{2\pi n}{L+1}\,.  
\end{equation}
Strictly speaking, this formula is valid within a limited range of $n$ and $L$, namely $n\leqslant 2$ and $L\leqslant 5$ because the operator mixing has been explicitly resolved on a case-by-case basis \cite{Georgiou:2012zj}. But later, when we re-derive this formula from the Bethe ansatz, we will see that it extrapolates  as written to arbitrary $n$ and $L$.

\paragraph{Dimension-four scalars.} Since  one-loop corrections for dimension-four operators were not calculated in \cite{Ivanovskiy:2024vel}, we need to perform the calculation from scratch. The two-loop mixing of the dimension-four singlets, necessary for their one-loop condensates, is described in  \cite{Georgiou:2012zj} where supersymmetry had been used to extract the fermionic components. However, two-loop Feynman diagrams are sufficiently simple in this case and we will recover the two-loop mixing pattern from a direct diagrammatic calculation, provising also an independent verification of the results in  \cite{Georgiou:2012zj}.
 
\subsection{Aside on dimension-four operators}
\label{subsec:dim4}
Lorentz singlets of dimension four are arguably the most important operators in any theory as they define the space of marginal couplings. It is thus worthwhile to make a digression on the dimension-four sector of the $\mathcal{N}=4$ SYM. 

We will allow ourselves to integrate by parts, by making no distinction between operators that differ by a total derivative. We also impose the equations of motion as they only affect contact terms in the correlation functions. Here is a typical identity that follows\footnote{We also used the fact that $C\Gamma ^i$ is a symmetric matrix and hence $\mathop{\mathrm{tr}}\bar{\Psi }\Gamma ^i\Psi\Phi _i=-\mathop{\mathrm{tr}}\bar{\Psi }\Gamma ^i\Phi _i\Psi $.}: 
\begin{equation}\label{IBPandEoM}
 -\mathop{\mathrm{tr}}D_\mu \Phi _iD^\mu \Phi _i\stackrel{{\rm IBP}}{= }
 \mathop{\mathrm{tr}}\Phi _iD^2\Phi _i
 \stackrel{{\rm EoM}}{=}\mathop{\mathrm{tr}}[\Phi _i,\Phi _j]^2+\mathop{\mathrm{tr}}\bar{\Psi }\Gamma ^i\Phi _i\Psi .
\end{equation}
Using this identity and the like any dimension-four Lorentz and R-symmetry singlet can be reduced to a linear combination of the following four operators\footnote{Mixing with double traces is large-$N$ suppressed \cite{Beisert:2003tq}.}:
\begin{equation}\label{O's}
 \mathcal{O}_1=\mathop{\mathrm{tr}}\Phi _i\Phi _i\Phi _j\Phi _j,
 \qquad 
 \mathcal{O}_2=\mathop{\mathrm{tr}}\Phi _i\Phi _j\Phi _i\Phi _j,
 \qquad 
 \mathcal{O}_3=\mathop{\mathrm{tr}}\bar{\Psi }\Gamma ^i\Phi _i\Psi ,
 \qquad 
 \mathcal{O}_4=\mathop{\mathrm{tr}}F_{\mu \nu }F^{\mu \nu },
\end{equation}
which we take as our basis. These operators are neither orthogonal nor equally normalized but being simple field monomials are perfectly suited for perturbative calculations. Correcting for orthonormality at every step, as often done in the literature, would be an unnecessary complication cluttering algebra without need.  

The dilatation operator in this sector is a $4\times 4$ matrix. In perturbation theory,
\begin{equation}\label{D-op}
 D=4+g^2H+g^4H'+\ldots 
\end{equation}
The $H_{ab}$ element of the mixing matrix is the coefficient of the $\mathcal{O}_a$ counterterm in the renormalization of $\mathcal{O}_b$. The easiest entry to understand (but the hardest to compute) is $H_{44}$ given by the diagrams of the pure glue theory, there defining the anomalous dimension of $\mathop{\mathrm{tr}}F_{\mu \nu }^2$. By the general principles of QFT the anomalous dimension of the Lagrangian density is the beta-function of the gauge coupling. The  direct calculation of $H_{44}$ however is rather subtle, giving rise to non-covariant counterterms in the most straightforward renormalization scheme \cite{Kluberg-Stern:1974iel}. Proper implementation of the background field method keeps gauge invariance at every step but requires to treat differently vertices with and without the external field \cite{Tarrach:1981bi,Grinstein:1988wz}. The final result nonetheless is simple, all the diagrams with internal loops cancel leaving behind the self-energy corrections \cite{Tarrach:1981bi}. The first two diagrams in the last line indeed reproduce the beta-function of QCD. It is important to take into  account extra vertices from the gauge-fixing term (\ref{gauge-fixing-BF}) to get the right result, while the scalar and fermion loops are the same as in the conventional Feynman gauge and agree with the divergent part of the self-energy that can be found for instance in  \cite{Erickson:2000af}. The diagrams in the last line mutually cancel giving $H_{44}=0$ which reflects the vanishing  of the  beta-function in the SYM.

\begin{figure}[t]
 \centerline{\includegraphics[width=15cm]{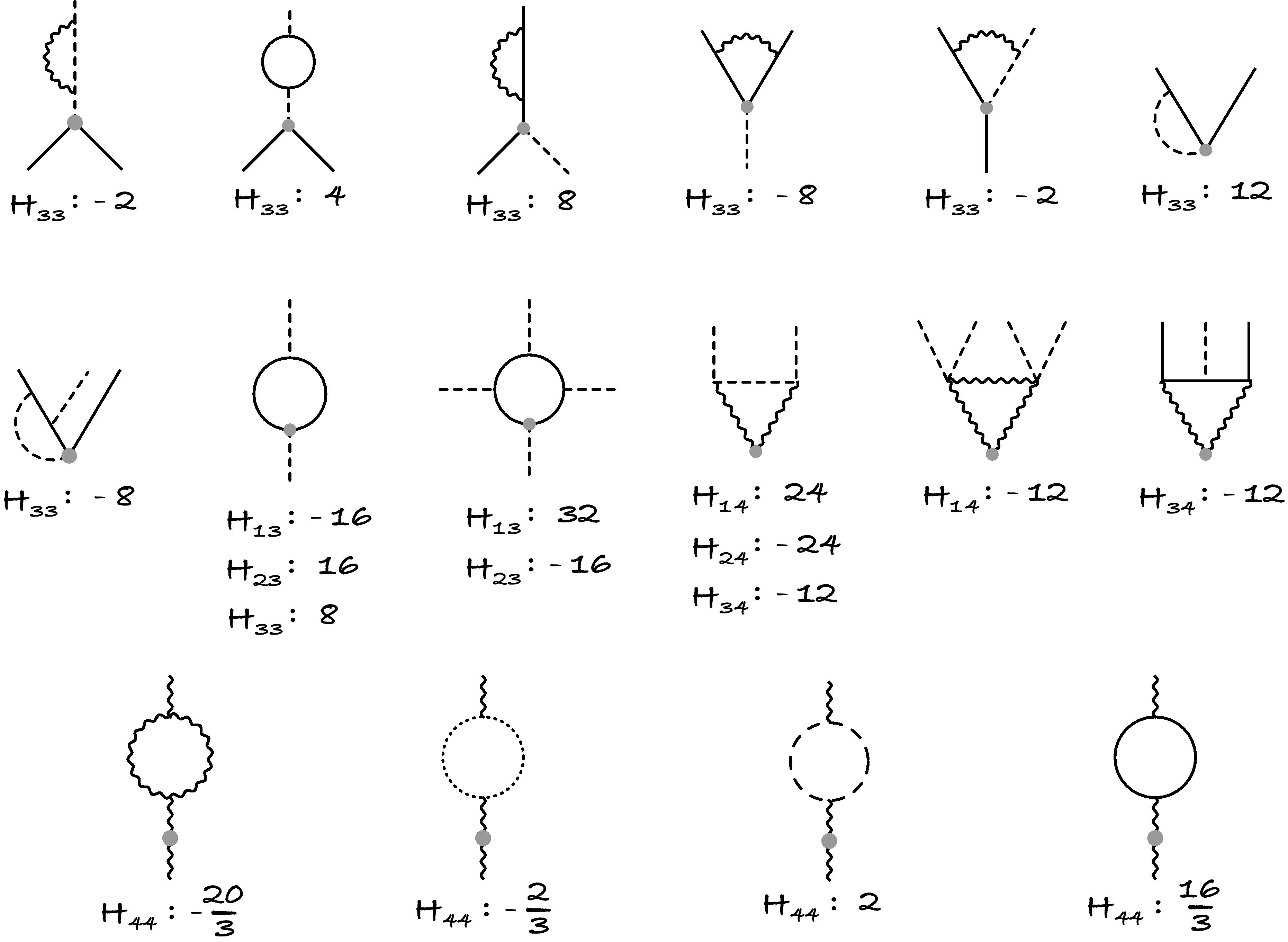}}
\caption{\label{H1}\small The one-loop counterterms for the operators (\ref{O's}).}
\end{figure}

The remaining matrix elements are computed from the diagrams in fig.~\ref{H1}.
We omit diagrams that identically cancel or do not diverge (e.g. pure glue internal loops in $H_{44}$) and do not display renormalization of the scalar operators $\mathcal{O}_1$ and $\mathcal{O}_2$, since  the corresponding matrix elements can be deduced from the $SO(6)$ dilatation operator (\ref{so6-Ham}). Collecting all pieces together we find:
\begin{equation}
 H=\begin{bmatrix}
 18  & -4  & 16 & 12 \\ 
  -4 & 8 & 0  & -24 \\ 
 0  & 0 & 12 & -24 \\ 
  0 & 0 & 0 & 0 \\ 
 \end{bmatrix}.
\end{equation}

The mixing matrix does not look Hermitian because the operator basis we are using is not orthonormal. This is not really a problem as long as we are interested in the eigenvalues and eigenvectors. What is more important is the upper-triangular structure, wherein the bosonic operators are regarded as a single block, that really simplifies the mixing pattern. The upper-triangular form is a consequence of disparity in length: the scalar operators have length four, the fermion operator $\mathcal{O}_3$ length three and the gluon operator $\mathcal{O}_4$ length two. One-loop diagrams cannot change length or, more precisely, lower-length operators can mix with higher-length ones but not vice versa. In general, a downgrade in length by $\ell$ units costs a factor of $\lambda ^{\ell+1}$ in the mixing matrix. At two loops, for example, the  scalar operators start mixing with fermions but not yet with gluons.

The eigenstates and eigenvalues of the one-loop mixing matrix are
\begin{align}\label{1-loop-ops}
 \mathcal{L}&=-\frac{1}{2}\,\mathcal{O}_4 +\mathcal{O}_1-\mathcal{O}_2-\mathcal{O}_3,\qquad \qquad \gamma _{\mathcal{L}}=0
\nonumber \\
\mathcal{O}_F&=\mathcal{O}_3+\frac{8}{5}(\mathcal{O}_2-\mathcal{O}_1),\qquad \qquad \gamma _F=12g^2
\nonumber \\
\mathcal{O}_\pm&=\mathcal{O}_1+\frac{5\mp\sqrt{41}}{4}\,\mathcal{O}_2,\qquad \qquad \gamma _\pm=\left(13\pm\sqrt{41}\right)g^2.
\end{align}
The first operator is the Lagrangian density:
\begin{equation}
 \mathcal{L}=-\frac{1}{2}\,\mathop{\mathrm{tr}}\left(F_{\mu \nu }^2
 +\left[\Phi _i,\Phi _j\right]^2+\bar{\Psi }\Gamma ^i\left[\Phi _i,\Psi \right]
 \right).
\end{equation}
The familiar form of the SYM Lagrangian (\ref{SYM-Lagrangian}) arises upon excluding fermions by their equations of motion and reshuffling bosonic terms with the help of the identity (\ref{IBPandEoM}). 

The second operator  belongs to the Konishi multiplet:
\begin{equation}
 \mathcal{O}_F=\mathop{\mathrm{tr}}\left(\frac{1}{2}\,\bar{\Psi }\Gamma ^i\left[\Phi _i,\Psi \right]+\frac{4}{5}\,\left[\Phi _i,\Phi _j\right]^2\right)=Q^4K,
\end{equation}
having the same anomalous dimension. We mention in passing that scalar fields enter $\mathcal{L}$ and $\mathcal{O}_F$ through commutators and hence the one-point functions of these operators vanish at tree level. And finally $\mathcal{O}_\pm$ are the two scalar operators (\ref{Opm}) we want to study.

\begin{figure}[t]
 \centerline{\includegraphics[width=6cm]{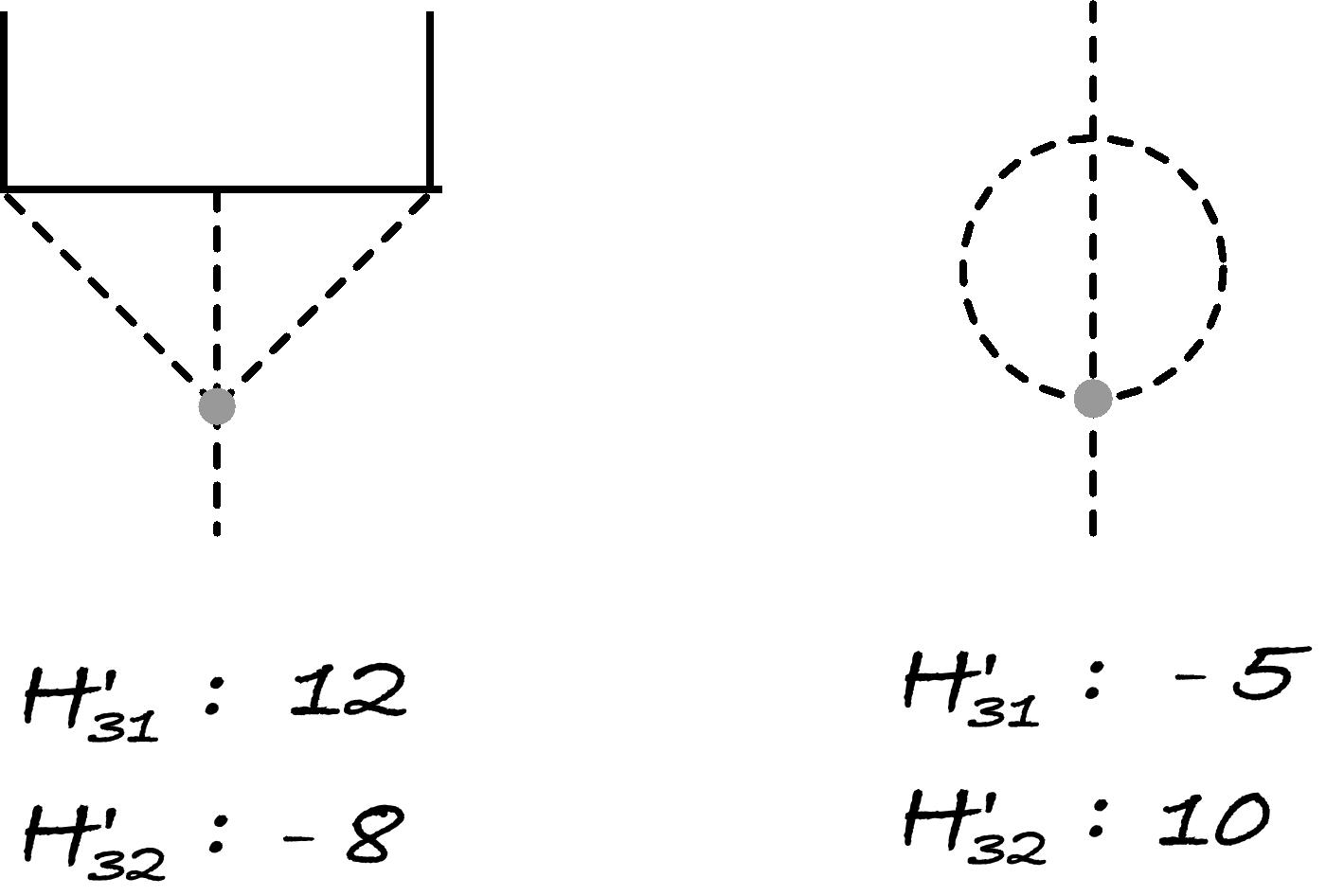}}
\caption{\label{H2}\small The two-loop $\mathcal{O}_{1,2}\rightarrow \mathcal{O}_3$ counterterms.}
\end{figure}

The upper-triangular form of the one-loop mixing greatly facilitates the two-loop analysis. As long as we interested in the scalar operators, the only matrix elements we need are $H'_{31}$, $H'_{32}$, and those are given by just two simple diagrams in fig.~\ref{H2}. We also need to understand how scalars mix among themselves, but scalar self-mixing is  determined by the two-loop $SO(6)$ Hamiltonian found in \cite{Georgiou:2011xj}:
\begin{align}
 \Gamma '=g^4\sum_{\ell=1}^{L}&\left(
 -8+12P_{\ell,\ell+1}-\frac{11}{2}\,K_{\ell,\ell+1}
 -2P_{\ell+1,\ell+2}P_{\ell,\ell+1}-2P_{\ell,\ell+1}P_{\ell+1,\ell+2}
 +K_{\ell,\ell+2}
 \right.
\nonumber \\
 &\left.
 +\frac{1}{2}\,P_{\ell+1,\ell+2}K_{\ell,\ell+2}
 +\frac{1}{2}\,P_{\ell,\ell+1}K_{\ell,\ell+2}
 +\frac{1}{2}\,P_{\ell+1,\ell+2}K_{\ell,\ell+1}
 +\frac{1}{2}\,P_{\ell,\ell+1}K_{\ell+1,\ell+2}
 \right.
\nonumber \\
 &\left.
 -\frac{1}{4}\,P_{\ell,\ell+2}K_{\ell,\ell+1}
 -\frac{1}{4}\,P_{\ell,\ell+2}K_{\ell+1,\ell+2}
 \right)
\end{align}
Applying this operator to $\mathcal{O}_1$ and $\mathcal{O}_2$ and collecting the results from fig.~\ref{H2} we get for the two-loop mixing matrix:
\begin{equation}\label{H'}
 H'=\begin{bmatrix}
 -96  & 32 & * & * \\ 
  34 & -4 & * & * \\ 
  7 & 2 & * & * \\ 
  0 & 0 & * & * \\ 
 \end{bmatrix},
\end{equation}
where $*$ stands for "any number".

These partial data is sufficient to determine two-loop corrections to the scalar operators. Indeed, by standard  perturbation theory,
\begin{equation}
 \left|\mathcal{O}_n\right\rangle=\left|n\right\rangle
 +g^2\sum_{m\neq n}^{}\left|m\right\rangle\,
 \frac{\left\langle m\right|H'\left|n\right\rangle}{\left\langle m\right.\!\left| m \right\rangle\left(E _n-E _m\right)}\,,\qquad 
 \qquad 
 \Delta _n=4+g^2E_n+g^4\,\frac{\left\langle n\right|H'\left|n\right\rangle}{\left\langle n\right.\!\left|n \right\rangle}\,.
\end{equation}
Since the mixing matrix is not Hermitian in the usual sense we need to distinguish left and right eigenvectors. The right eigenvectors of $H$ are the linear combinations in (\ref{1-loop-ops}) and the left ones are
\begin{align}\label{left-values}
 \left\langle \mathcal{L}\right|&=\begin{bmatrix}
  0 & 0 & 0 & 1 \\ 
 \end{bmatrix}
\nonumber \\
\left\langle F\right|&=
\begin{bmatrix}
  0 & 0  & 1  &  -2  \\ 
 \end{bmatrix}
\nonumber \\
\left\langle \pm\right|&=
\begin{bmatrix}
  1 & \frac{5\mp\sqrt{41}}{4} & \frac{-2\pm 2\sqrt{41}}{5} & \frac{3\mp3\sqrt{41}}{10} \\ 
 \end{bmatrix}.
\end{align}
Crucially, all overlaps between $\left\langle \mathcal{L}\right|$ and $\left|\pm\right\rangle$ vanish, and the two-loop corrected scalar operators are completely determined by (\ref{1-loop-ops}), (\ref{H'}) and (\ref{left-values}), without the need to know $H'$ completely:
\begin{equation}
 \widehat{\mathcal{O}}_\pm=\mathcal{O}_\pm-\frac{127\mp 13\sqrt{41}}{205}\,g^2\mathcal{O}_\mp-\frac{3\mp\sqrt{41}}{4}\,g^2\mathcal{O}_F.
\end{equation}
The eigenvalues are readily computed as well and fix the two-loop anomalous dimensions to be
\beq\label{eq:twist4delta}
\Delta_{\pm} = 4 + \left(13\pm \sqrt{41}\right)g^2 -\left(46\pm \frac{310}{\sqrt{41}}\right)g^4 +\mathcal{O}(g^6)\period
\eeq
We will reproduce this result from the Bethe ansatz in the next section.

The one-loop one-point functions (\ref{1-loop-gen}) are expressed through the tree-level data, the operator norm and the vacuum expectation value (we omit the subscript ``tree" in what follows). The correlation matrix of the basis (\ref{O's}) can be readily computed, by simply counting the tree-level contractions (we omit $\mathcal{O}_4$ since it does not appear in $\widehat{\mathcal{O}}_\pm$):
\begin{equation}
 \left\langle \mathcal{O}_a\,\mathcal{O}_b\right\rangle
 =\begin{bmatrix}
  21  & 6  & 0 \\ 
  6 & 36 &  0 \\ 
  0 & 0 & \frac{48}{g^2} \\ 
 \end{bmatrix}.
\end{equation}
The fermion operator $\mathcal{O}_3$ has a bigger norm because of the smaller length.  Its two-point function lacks one propagator compared to  scalars\footnote{We use the length-four normalization for the norm in (\ref{2pt-tree}) taking out a factor of $4(2g^2)^4$, and also for the fermion operator even if it has length three. The overall normalization is just a convention but it should be consistently applied to all operators at hand.}.  

After a simple algebra we find for the norm of the two-loop eigenstates:
\begin{equation}
 \left\langle \widehat{\mathcal{O}}_\pm\,\widehat{\mathcal{O}}_\pm \right\rangle
 =\frac{3}{2}\left(123\mp 17\sqrt{41}\right)-6\left(25\mp 3\sqrt{41}\right)g^2+\mathcal{O}\left(g^4\right).
\end{equation}
As a cross-check  we can also verify that the two operators are orthogonal to the desired accuracy:
\begin{equation}
 \left\langle \widehat{\mathcal{O}}_+\,\widehat{\mathcal{O}}_- \right\rangle
 =0+0\cdot g^2+\mathcal{O}\left(g^4\right).
\end{equation}
 One may worry that mixing with gluons, whose norm is enhanced by $1/g^4$, may also contribute, but an admixture of $\mathcal{O}_4$ into $\mathcal{O}_\pm$ starts at three loops and thus shifts the norm by $(g^4)^2/g^4=g^4$.
 
The tree-level expectation value of (\ref{Opm}) is obtained by simply replacing all $\Phi _i$'s by $v$:
\begin{equation}
 \left\langle \mathcal{O}_\pm\right\rangle=\frac{9\mp\sqrt{41}}{4}\,v^4,
\end{equation}
 and hence
\begin{equation}
  \left\langle \widehat{\mathcal{O}}_\pm\right\rangle=\frac{9\mp\sqrt{41}}{4}\left(1-\frac{59\pm 7\sqrt{41}}{82}\,g^2\right)v^4,
\end{equation}
where we have taken into account that $\left\langle \mathcal{O}_F\right\rangle=0$.

Collecting the pieces together we get from (\ref{1-loop-gen}):
\begin{equation}\label{eq:twist4fieldtheory}
  \left\langle  \widehat{\mathcal{O}}_\pm\right\rangle_{\rm 1-loop}
  =\frac{v^4}{32g^4}\,\sqrt{\frac{1}{2}\mp\frac{7}{6\sqrt{41}}}\left\{
  1+
  \left[
  (13\pm\sqrt{41})\left(\gamma _E+\ln\frac{v}{2}\right)
  -\frac{49\pm 21\sqrt{41}}{410}
  \right]g^2
  \right\}.
\end{equation}
Here again we find a universal term $\ln(v\,{\rm e}\,^{\gamma _E}/2)$ multiplying the one-loop anomalous dimension and an operator-specific correction due to the norm arising from the two-loop operator mixing.

\subsection{Konishi and two-magnon operators at one loop}\label{sec:KonishiOneLoop}

Here we extend to one-loop the tree-level one-point functions of two-magnon operators given in section \ref{sec-examples}. 

The simplest non-protected operator with a non-zero one-point function is the Konishi operator. This is a twist-2 single-trace and R-charge singlet operator of dimension:
\beq
\Delta_K = 2 \,+\, 12 g^2 - 48 g^4 +\mathcal{O}(g^6) 
\eeq
% and a pair of Bethe roots encoded in the polynomial:
% \beq
% \prod_{i=1}^{2}(z-u_{4,i}) = z^2 -\frac{1}{12}-\frac{4}{3}g^2 +\mathcal{O}(g^4)
% \eeq
In the $SU(2)$ grading the corresponding supermultiplet has occupation numbers: 
$\bar{\mathbf{K}}\,=\,\{0,0,1,2,1,0,0\}$.
The pair of middle-node roots are given by:
\beq\label{eq:uLnroot}
u_{4,1}=-u_{4,2} = u = \frac{1}{2}\cot\left(\frac{n\pi}{L+1}\right)+2g^{2}\frac{L+2}{L+1}\sin\left(\frac{2 n\pi}{L+1}\right)+\mathcal{O}(g^4)
\eeq
For the Konishi operator we should set $L=2$ and $n=1$. For more general $L$ and $n$ these roots correspond to the two-magnon operators  introduced in \eqref{Konishi-nL}.

For this family of operators, the unpaired fermionic Bethe roots take the special value: $x(u_3)=x(u_5)=0$ at all loops (only at leading order this is equivalent to $u_3=u_5=0$). At this value the elements of the Gaudin matrix with $\partial_{u_3}$ become zero leading to a zero determinant. However, in order to obtain a non-zero result, we prescribe to do the change: $\partial_{u_{3,5}} \to \frac{1}{g}\partial_{x(u_{3,5})}$ for these fermionic roots.  Following this prescription we obtain the matrices $G^{\pm}$ as:
\beq
G^{+} = \left(\begin{array}{ccc}
 \frac{1}{g}\frac{\partial \chi_{3}(u_3)}{\partial x(u_3)}    & \frac{\partial \chi_{3}(u_3)}{\partial u_{4,1}}+\frac{\partial \chi_{3}(u_3)}{\partial u_{4,2}} & 0  %\frac{\partial \chi_{3}(u_3)}{\partial x(u_5)}
 \\
  \frac{1}{g}\frac{\partial \chi_{4}(u_{4,1})}{\partial x(u_3)}     & \frac{\partial \chi_{4}(u_{4,1})}{\partial u_{4,1}}+\frac{\partial \chi_{4}(u_{4,1})}{\partial u_{4,2}} & \frac{1}{g}\frac{\partial \chi_{4}(u_{4,1})}{\partial x(u_5)}  \\
  0 %\frac{\partial \chi_{5}(u_5)}{\partial x(u_3)}
  & \frac{\partial \chi_{5}(u_{5})}{\partial u_{4,1}}+\frac{\partial \chi_{5}(u_{5})}{\partial u_{4,2}} & \frac{1}{g}\frac{\partial \chi_{5}(u_5)}{\partial x(u_5)}
\end{array}\right) \quad\text{and}\quad G^{-} = \frac{\partial \chi_{4}(u_{4,1})}{\partial u_{4,1}}-\frac{\partial \chi_{4}(u_{4,1})}{\partial u_{4,2}}
\eeq
where we omit the unpaired roots $u_{3,5}$ in the definition of $G^{-}$, as described below \eqref{eq:defGpm}, resulting in a rank-one matrix.  The functions $\chi$ are defined in \eqref{eq:su2BetheEq} and here they specialize to:
\begin{align}
e^{i\chi_4(z)} &= \left[\frac{x^{+}(z)}{x^{-}(z)}\right]^{L}\,\prod_{j=1}^{2} \frac{z-u_{4,j}-i}{z-u_{4,j}+i}\frac{1}{\sigma^2(z,u_{4,j})}\times \frac{x^{+}(z)-x(u_3)}{x^{-}(z)-x(u_3)}\,\frac{x^{+}(z)-x(u_5)}{x^{-}(z)-x(u_5)} \\
e^{i\chi_3(z)}=e^{i\chi_5(z)} &= \prod_{j=1}^{2}\frac{x(z)-x^{+}(u_{4,j})}{x(z)-x^{-}(u_{4,j})}
\end{align}
%%%%%%%%%%%%%
The corresponding matrices $G^{\pm}$ can be written using the auxiliary functions:
\begin{align}
    \mathbb{K}&\equiv -i\partial_v \log\frac{x^{+}(v)}{x^{-}(v)}\bigg{|}_{v\to u} = -\frac{1}{u^2+\frac{1}{4}}-2g^2\frac{3\,u^2-\frac{1}{4}}{\left(u^2+\frac{1}{4}\right)^3} + \mathcal{O}(g^4) \\ %\quad \text{and} \quad
    \mathbb{P}&\equiv  \frac{-i}{g}\partial_{x(v)} \log\frac{x(v)-x^{+}(u)}{x(v)-x^{-}(u)}\bigg{|}_{x(v)\to 0} = \frac{1}{u^2+\frac{1}{4}}+g^2\frac{3\,u^2-\frac{1}{4}}{\left(u^2+\frac{1}{4}\right)^3} + \mathcal{O}(g^4)
\end{align}
Then the superdeterminant evaluates to:
\begin{align}
{\rm SdetG} =  \frac{\det G^{+}}{\det G^{-}} = \frac{\det\left(\begin{array}{ccc}
        2\mathbb{P} & 2 \mathbb{K} & 0  \\
        \mathbb{P} & (L+2) \mathbb{K} & \mathbb{P}  \\
        0 & 2\mathbb{K} & 2\mathbb{P} 
    \end{array} \right) }{(L+2)\mathbb{K} + \frac{1}{u^2+\frac{1}{4}}} = \frac{4L\,\mathbb{K}\,\mathbb{P}^2}{(L+2)\mathbb{K} + \frac{1}{u^2+\frac{1}{4}}} 
\end{align}
The other factors entering the formula \eqref{exact-1pt2} are given by\footnote{We do not include the unpaired fermionic roots $x(u_{3,k})=x(u_{5,k})=0$ in the denominator of \eqref{exact-1pt2}, in order  to obtain a finite result.}:
\beq
\prod_{k=1}^{2}\sigma_{B}(u_{4,k})\,\frac{u_{4,k}\left(u_{4,k}+\frac{i}{2}\right)}{\sigma(u_{4,k},\bar{u}_{4,k})}\,=\,\left(\frac{v}{2g}\right)^4 \left[u^{2}(u^2+1/4) + 8\,g^{2}\,\log(v/2 e^{\gamma_E}) + \mathcal{O}(g^4)\right]
\eeq
and the symmetry factor in \eqref{eq:CKn1} evaluates to:
\beq
\mathbb{C}_{\textbf{K}}\,=\frac{1}{1+\gamma/2} =\, 1-\frac{2g^{2}}{u^2+\frac{1}{4}} +\mathcal{O}(g^4)
\eeq
Finally the integrability prediction gives:
\beq
\langle \mathcal{O}_{L,n} \rangle_{1\text{-loop}}=\mathbb{C}_{\textbf{K}}\times
\frac{1}{\sqrt{L}}\left(\frac{v}{2g}\right)^{L-2}\,\sqrt{\prod_{k=1}^{2}\sigma_{B}(u_{4,k})\,\frac{u_{4,k}\left(u_{4,k}+\frac{i}{2}\right)}{\sigma(u_{4,k},\bar{u}_{4,k})}\times {\rm SdetG}
}
\eeq
and by using the one-loop series of $u$ in \eqref{eq:uLnroot} we reproduce the field-theory result in \eqref{av-int}. 

% Using these Bethe roots we compute the superdeterminant: 
% \beq
% \frac{\det G^{+}}{\det G^{-}} = \# + \mathcal{O}(g^4) 
% \eeq
% the prefactors:
% \beq
% \prod\limits_{j=1}^{2} u_{4,j}\bigg{(}u_{4,j}-\frac{i}{2}\bigg{)} \,=\,\frac{1}{36} + \frac{5g^2}{9}+\mathcal{O}(g^4) \quad \text{and}\quad \prod\limits_{j=1}^{2}\sigma_{B}(u_{4,j}) \,=\, 1 + 24 \log(v/2 e^{\gamma_E}) g^2 + \mathcal{O}(g^4) 
% \eeq
% and the symmetry factor: 
% \beq
% \mathbb{C}_{\mathbf{n}\nu}^{\text{twist-}2} \,=\, 1-\sum_{j=1}^{2} \frac{g^2}{u_{4,j}^2+\frac{1}{4}} + \mathcal{O}(g^4) \,=\,  1-6g^2 +\mathcal{O}(g^4)
% \eeq

\subsection{Dimension-4 operators at one loop}\label{sec:twist4oneloop}
We next study the dimension-4 operators discussed in section \ref{subsec:dim4}. These are $SO(6)$ R-charge singlets, at leading order they are given by the linear combination of color traces in \eqref{Opm} and their scaling dimensions $\Delta_{\pm}$ are given in \eqref{eq:twist4delta}.

These operators and their one-point functions were presented in section \ref{sec-examples} at tree level using the $SO(6)$-grading formulation of the overlap in \eqref{SO(6)-overlap} and \eqref{rootsOpm}. Instead, here we use the all-loop formula \eqref{exact-1pt} in the $SU(2)$ grading and specialize to one loop. The tree-level limit of this latter is given by \eqref{eq:su2treelevel}.

In the $SU(2)$ grading, the corresponding superconformal multiplets are characterized by occupation numbers $\bar{\mathbf{K}} = \{0,1,2,4,2,1,0\}$ and by the Bethe roots encoded in the polynomials: 
\begin{align}
\prod_{i=1}^{4}(z-u_{4,i}) &= z^4 + z^2\left(-\frac{1}{10}\mp\frac{\sqrt{41}}{15}+g^2\left(-\frac{322}{75}\mp\frac{1238}{75\sqrt{41}}\right)+\mathcal{O}(g^4)\right) \nonumber\\
&\qquad+\, \left(\frac{17}{240}\mp\frac{\sqrt{41}}{120}+g^2\left(\frac{19}{50}\pm\frac{203}{150\sqrt{41}}\right)+\mathcal{O}(g^4)\right)\\
\prod_{i=1}^{2}(z-u_{3,i})&=\prod_{i=1}^{2}(z-u_{5,i})=z^2 + \frac{1}{20}(1\mp\sqrt{41}) +g^2 \left(-\frac{78}{25}\mp\frac{412}{25\sqrt{41}}\right) +\mathcal{O}(g^4)\\
\prod_{i=1}^{1}(z-u_{2,i})&=\prod_{i=1}^{1}(z-u_{6,i})= z
\end{align}
This corresponds to Bethe solutions with symmetric wings. The set of roots $u_{3,j}, u_{4,j}$ and $u_{5,j}$ come in pairs of the type $\{u,-u\}$. On the other hand, the roots at the bosonic nodes $2$ and $6$ are unpaired and are exact at zero at all loops: $u_{2,1}=u_{6,1}=0$.

By using the formula \eqref{exact-1pt} we can compute the corresponding one-point function. With this purpose, we start by computing the superdeterminant. As an example, below we show the matrices $G^{\pm}$, following def. \eqref{eq:defGpm}, for the operator with dimension $\Delta_{+}$ in \eqref{eq:twist4delta}. We organize the rows and columns of these matrices according to the six roots:
\beq
\left\{u_{2,1}=0,\,u_{3,1}= -0.51 i -0.11 i\,g^2,\,u_{4,1}= -0.70-4.36\,g^2,\,u_{4,2}= -0.19-2.0\,g^2,\,u_{5,1},\,u_{6,1}\right\}
\eeq
with $u_{5,1}=u_{3,1}$ and $u_{6,1}=u_{2,1}$, while the  missing roots correspond to the parity partners ($u_{4,3}=-u_{4,1}\,,\cdots$). With this organization the matrices $G^{\pm}$ for the operator with dimension $\Delta_{+}$ are given by:
\begin{align}
G^{+}&={\scriptsize\left(
\begin{array}{cccccc}
 -3.84+42.1\,g^2 & 3.84-42.1\,g^2 & 0 & 0 & 0 & 0 \\
 1.92-21.0 \,g^2 & 6.30+17.6 \,g^2 & 4.11+7.12 \,g^2 & 4.11-10.57 \,g^2 & 0 & 0 \\
 0 & -4.11+2.86\, g^2 & -10.92+16.1\, g^2 & -2.70+10.08 \,g^2 & -4.11+2.86 \,g^2 & 0 \\
 0 & -4.11+36.5 \,g^2 & -2.70+10.08\, g^2 & -19.5+149. \,g^2 & -4.11+36.5\, g^2 & 0 \\
 0 & 0 & 4.11+7.12 \,g^2 & 4.11-10.57\, g^2 & 6.30+17.6\, g^2 & 1.92-21.0 \,g^2 \\
 0 & 0 & 0 & 0 & 3.84-42.1\, g^2 & -3.84+42.1\, g^2 \\
\end{array}
\right)
} \\
G^{-}&={\scriptsize\left(
\begin{array}{cccc}
 6.30+17.6\,g^2 & 2.96+13.0\,g^2 & 1.45+1.082\, g^2 & 0 \\
 -2.96-3.18\,g^2 & -9.57+4.96\,g^2 & -0.468-4.06\,g^2 & -2.96-3.18\,g^2 \\
 -1.45+16.7\,g^2 & -0.468-4.06\,g^2 & -16.0+140.2\,g^2 & -1.45+16.7\,g^2 \\
 0 & 2.96+13.0\,g^2 & 1.45+1.082\,g^2 & 6.30+17.6\,g^2 \\
\end{array}
\right)}
\end{align}
where we only show low numerical precision to save space and define the matrix $G^{-}$ omitting the zero roots $u_{2,1}$ and $u_{6,1}$, as described below eq.~\eqref{eq:defGpm}. By using these matrices, we compute the superdeterminant to order $\mathcal{O}(g^2)$ but numerically exact: 
\beq
\frac{\det G^{+}}{\det G^{-}} = \frac{96(287\mp 3\sqrt{41})}{1025}\left(1+g^{2}\left(-\frac{3083}{205}\mp\frac{83}{\sqrt{41}}\right)+\mathcal{O}(g^4)\right)
\eeq
On the right-hand side the options $\mp$ correspond to dimensions $\Delta_{\pm}$ respectively.  

The other ingredients in \eqref{exact-1pt2} are the Q-function prefactor: 
\beq
\frac{\prod\limits_{j=1}^{4} \frac{u_{4,j}\left(u_{4,j}+\frac{i}{2}\right)}{\sigma(u_{4,j},\bar{u}_{4,j})}}{\prod\limits_{j=1}^{2}x(u_{3,j})\,\prod\limits_{j=1}^{2}x(u_{5,j})}\,=\, \frac{21\mp\sqrt{41}}{288} +g^{2}\frac{1}{72}\left(29\pm\frac{221}{\sqrt{41}}\right)+\mathcal{O}(g^4)
\eeq
as well as the boundary dressing-phase factor from \eqref{weak-dressing}: 
\beq
\prod\limits_{j=1}^{4} \sigma_{B}\left(u_{4,j}\right) =\left(\frac{v}{16}\right)^{4} \left[ 1+2g^2(13\pm\sqrt{41})\log(v/2\,e^{\gamma_E}) \,+\, \mathcal{O}(g^4)\right]
\eeq
and the symmetry factor, from \eqref{eq:CKn2} with $\mathbb{K}_2=\mathbb{K}_6= \frac{\Delta_{\pm}-2}{2}$, evaluates to:
\beq
\mathbb{C}_{\mathbf{K}}^{\pm} \,=\,  1-\frac{2}{\Delta_{\pm}}\,=\,  \frac{1}{2}+g^{2}\frac{13\pm\sqrt{41}}{8} +\mathcal{O}(g^4)
\eeq
Assembling all together, according to \eqref{exact-1pt2}, we obtain the integrability prediction for the one-point function:
\beq
\left\langle  \widehat{\mathcal{O}}_\pm\right\rangle
  =\frac{v^4}{32g^4}\,\sqrt{\frac{1}{2}\mp\frac{7}{6\sqrt{41}}}\left\{
  1+
  \left[
  (13\pm\sqrt{41})\left(\gamma _E+\ln\frac{v}{2}\right)
  -\frac{49\pm 21\sqrt{41}}{410}
  \right]g^2
  \right\}
\eeq
which perfectly matches with the field-theory computation in \eqref{eq:twist4fieldtheory}.

    \section{Conclusion}\label{sec:conclusion}
    In this paper, we studied vacuum condensates on the Coulomb branch of planar $\mathcal{N}=4$ SYM. By formulating and solving integrable bootstrap equations, we determined the boundary state corresponding to the probe D3-brane in AdS. We then used it to conjecture the asymptotic formula for the vacuum condensates at finite 't Hooft coupling. We tested the formula at tree level and one-loop, finding perfect agreement with the field theory computation. 
    
    Our results provide direct and constructive evidence for integrability on the Coulomb branch and motivate further study on the subject. 
    
    First, our asymptotic formula contains the kinematical factor $\mathbbm{C}_{\mathbf{K}}$, needed for the superconformal descendants. Although well-motivated by the appearance of similar-looking factors in related contexts \cite{Gombor:2024api,Bercini:2022jxo}, we currently lack a rigorous derivation beyond tree level and it is important to perform further tests or come up with a proof. It is also important to extend our conjecture for the kinematical factor to a more general class of operators. 
    
    Second, it is important to establish integrability on the Coulomb branch on the string theory side, i.e.~classical integrability for the boundary condition corresponding to the probe D3 brane \cite{upcoming}. 

    Third, it would be interesting to perform the integrability analysis of other observables on the Coulomb branch beyond the vacuum condensates, for instance the form factors\footnote{The form factors of $\mathcal{N}=4$ SYM at the conformal point have been studied from integrability in \cite{Sever:2020jjx,Basso:2023bwv}.} and three- (and four-)point scattering amplitudes of W-boson bound states. For this purpose, it may be useful to revisit the spectrum of the W-boson bound states---which was determined indirectly using the Regge theory \cite{Caron-Huot:2014gia}---with the standard integrability approaches; e.g.~  the classical integrability of the string sigma model and the spin chain picture at weak coupling. These W-boson bound states appear also as intermediate states in four-point amplitudes studied in \cite{Alday:2009zm, Arkani-Hamed:2023epq, MdAbhishek:2023nvg, Flieger:2025ekn, He:2025vqt, Cortes:2025xmu} at different points of the Coulomb-branch moduli. In addition, there is a conjecture that relates these massive four-point amplitudes with four-point correlators of large R-charge operators at the conformal point \cite{Caron-Huot:2021usw}. It would be enlightening to revisit these results from the point of view of integrability on the Coulomb branch.

    Finally, the study of the Coulomb branch of $\mathcal{N}=4$ SYM is the first step towards applying integrability beyond conformal field theory. It would be interesting to study other integrable setups where the conformal symmetry is broken, either explicitly or spontaneously. Potential candidates include the massive fishnet theory \cite{Loebbert:2020tje,Loebbert:2020hxk}, the fishnet theory on the moduli space \cite{Karananas:2019fox} and the mass-deformed ABJM theory \cite{Gomis:2008vc,Kristjansen:2024zvl}.  
    \subsection*{Acknowledgment} We thank Clay C\'{o}rdova for the collaboration at the early stage of this project. The work of F.C. is supported in part by the Simons Foundation
grant 994306 (Simons Collaboration on Confinement and QCD Strings), as well the NCCR
SwissMAP that is also funded by the Swiss National Science Foundation.
The work of K.~Z. was supported by VR grant 2021-04578. 

\appendix
\section{Coulomb branch from defect one-point function}
In this appendix, we explain details on how to take the limit of the results for the defect one-point function \cite{Komatsu:2020sup, Gombor:2020auk, Gombor:2021uxz} to get the results for the Coulomb branch.
\subsection{Boundary dressing phase}\label{ap:sigmaB}
 The solution to the crossing equation for the defect one-point function $\sigma_B^{D}$ is given by \cite{Komatsu:2020sup}
\beq
\sigma_B^{D}(u)=2^{-4E(u)}\frac{G(x^{-},x_s)G(x^{+},-1/x_s)}{G(x^{+},x_s)G(x^{-},-1/x_s)}\comma
\eeq
where the function $G(x,y)$ for $|x|>1$ and $|y|>1$  is given by a contour integral
\beq
\frac{1}{i}\log G(x,y)=\frac{2}{i}\oint_{|z|=1}\frac{dz}{2\pi i}\oint_{|w|=1}\frac{dw}{2\pi i}\frac{1}{x-z}\frac{1}{y-w}\log \mathfrak{G}(z,w)\comma
\eeq 
with
\beq
\mathfrak{G}(z,w)=\frac{\Gamma \left[1+ig (z+\tfrac{1}{z}+w+\tfrac{1}{w})\right]}{\Gamma \left[1-ig (z+\tfrac{1}{z}-w-\tfrac{1}{w})\right]}\period
\eeq
In other parameter regimes (of $x$ and $y$), $G(x,y)$ is given by an analytic continuation of the expression above. 
As is clear from this integral representation, $G(x,x_s)\to 1$ in the limit $x_s\to \infty$. Therefore, what we need to study  is the limiting behavior of $G(x,-1/x_s)$. 

As explained in section 4.5 of \cite{Komatsu:2020sup}, performing the analytic continuation and being careful about the contribution from a pole crossing the integration contour, we get the following integral representation for $G(x,1/y)$.
\beq\label{Gx1/y}
\frac{1}{i}\log G(x,1/y)=\chi_{\rm int}(x,y) +\chi_{\rm pole}(x,y)\comma
\eeq
with
\beq
\begin{aligned}
&\chi_{\rm int} (x,y)=\frac{2}{i}\oint_{|z|=1}\frac{dz}{2\pi i}\oint_{|w|=1}\frac{dw}{2\pi i}\frac{1}{x-z}\frac{1}{\frac{1}{y}-w}\log \mathfrak{G}(z,w)\comma\\
&\chi_{\rm pole}(x,y)=\frac{2}{i}\oint_{|z|=1}\frac{dz}{2\pi i}\frac{1}{x-z}\log \mathfrak{G}(z,y)\period
\end{aligned}
\eeq
Now, clearly $\chi_{\rm int}$ remains finite in the limit $y\to \infty$:
\beq
\tilde{\chi}(x)\equiv \chi_{\rm int}(x,\infty)=-\frac{2}{i}\oint_{|z|=1}\frac{dz}{2\pi i}\oint_{|w|=1}\frac{dw}{2\pi i}\frac{1}{x-z}\frac{1}{w}\log \mathfrak{G}(z,w)\period
\eeq
On the other hand, expanding $\log \mathfrak{G}(z,y)$ around $y\to \infty$ using Stirling's formula, we get
\beq
\begin{aligned}
\chi_{\rm pole}(x,y)&=\frac{2}{i}\oint_{|z|=1}\frac{dz}{2\pi i}\frac{2i g}{x-z}\left(z+\frac{1}{z}\right)\log (i gy)+O(1/y)\\
&=\frac{4g}{x}\log (i gy)+O(1/y)
\end{aligned}
\eeq 
As a result, we get
\beq
\sigma_B^{D}(u)=2^{-4E(u)}\exp\left[4ig\left(\frac{1}{x^{+}}-\frac{1}{x^{-}}\right)\log (- i gx_s)\right]e^{i(\tilde{\chi}(x^{+})-\tilde{\chi}(x^{-}))}\left(1+O(1/x_s)\right)\period
\eeq
Combining this with the prefactor $x_s^2$ in \eqref{eq:limittotake}, we get
\beq\label{eq:solution0}
x_s^{2}\sigma_B^{D}(u)=-\frac{2^{-4E(u)}}{g^2}\exp\left[\left(2+4ig\left(\frac{1}{x^{+}}-\frac{1}{x^{-}}\right)\right)\log (- i gx_s)\right]e^{i(\tilde{\chi}(x^{+})-\tilde{\chi}(x^{-}))}\left(1+O(1/x_s)\right)\period
\eeq 
This basically shows that the naive limit is actually ill-defined owing to the factor $\log (-i gx_s)$. However, there is a simple way to remove this divergence: the crucial observation is that the prefactor multiplying $\log (-2i x_s)$ is given precisely by the energy of a magnon\fn{The first equality is a little nontrivial to check; we need to use the relation
\beq
g(x^{+}-x^{-})(1-\tfrac{1}{x^{+}x^{-}})=g(x^{+}+1/x^{+})-g(x^{-}+1/x^{-})=i\period
\eeq
} $E(u)$
\beq
2+4i g \left(\frac{1}{x^{+}}-\frac{1}{x^{-}}\right)=2\frac{1+\frac{1}{x^{+}x^{-}}}{1-\frac{1}{x^{+}x^{-}}}=4 E(u)\period
\eeq
We can therefore rewrite \eqref{eq:solution0} as
\beq
x_s^{2}\sigma_B^{D}(u)=-\frac{1}{g^2}\left(\frac{- i gx_s}{2}\right)^{4E(u)}e^{i(\tilde{\chi}(x^{+})-\tilde{\chi}(x^{-}))}\left(1+O(1/x_s)\right)\period
\eeq
We then use the fact that the crossing equation admits a so-called CDD ambiguity (see the explanation below (4.51) in \cite{Komatsu:2020sup}); Namely if we are given a solution to the crossing equation, one can construct infinitely many new solutions by multiplying $\pm e^{f_{\rm odd}(E)}$ where $f_{\rm odd}$ is an odd function of the energy $E(u)$. in this case, by choosing the CDD factor to be\fn{Here we chose the CDD factor so that it gives a finite answer in the limit and reproduces the tree-level result.}
\beq
\sigma_{\rm CDD}(u)=-\left(\frac{v}{-i gx_s}\right)^{4E(u)}\comma
\eeq
we obtain
\beq
\sigma_{\rm CDD}x_s^{2}\sigma_{B}^{D}(u)=\frac{\left(v/2\right)^{4E(u)}}{g^2}e^{i(\tilde{\chi}(x^{+})-\tilde{\chi}(x^{-}))}(1+O(1/x_s))\period
\eeq
After this manipulation, we can safely take the limit $x_s\to \infty$. The result reads
\beq
\sigma_B(u)\equiv \lim_{x_s\to \infty}\sigma_{\rm CDD}x_s^2 \sigma_B^{D}(u)=\frac{(v/2)^{4E(u)}}{g^2}e^{i(\tilde{\chi}(x^{+})-\tilde{\chi}(x^{-}))} \comma
\eeq
with
\beq
\tilde{\chi}(x)\equiv-\frac{2}{i}\oint_{|z|=1}\frac{dz}{2\pi i}\oint_{|w|=1}\frac{dw}{2\pi i}\frac{1}{x-z}\frac{1}{w}\log \mathfrak{G}(z,w)\period
\eeq

\subsection{Crossing equation}\label{appendix:solve-srossing}

We can also check directly  that the dressing phase (\ref{sigmaB-final}) solves the crossing equation (\ref{eq:crossingtosolve}). Crossing acts on $x^\pm$ as
\begin{equation}
 x^\pm(u^{2\gamma })=\frac{1}{x^\pm(u)}\,,
\end{equation}
and it is easy to see that $E(u^{2\gamma })=-E(u)$. The CDD factor therefore solves the homogeneous equation. The rest reads
\begin{equation}\label{rest-of-crossing}
 \frac{1}{g^4}\,\,{\rm e}\,^{i\tilde{\chi}(x^+)+i\tilde{\chi}(1/x^+)-i\tilde{\chi}(x^-)-i\tilde{\chi}(1/x^-)}=\left(\frac{x^+}{x^-}\right)^2.
\end{equation}

The crossing-transformed function $\tilde{\chi}(1/x)$ should be understood in terms of analytic continuation:
\begin{equation}
 \tilde{\chi}(1/x)=-\frac{2}{i}\oint\frac{dz}{2\pi i}\,\,\frac{1}{1/x-z}\oint
 \frac{dw}{2\pi iw}\,\log \mathfrak{G}(z,w)
 -\frac{2}{i} \oint\frac{dw}{2\pi iw}\,\log \mathfrak{G}(1/x,w).
\end{equation}
The second term cancels the pole inside the contour of integration, because the original contour should have left $1/x$ outside, cf.~(\ref{Gx1/y}).

Changing the integration variable in the first  term to $1/z$, we find:
\begin{equation}
  \tilde{\chi}(1/x)+\tilde{\chi}(x)=\frac{2}{i}\oint\frac{dz}{2\pi iz}
  \oint
 \frac{dw}{2\pi iw}\,\log \mathfrak{G}(z,w)
 -\frac{2}{i} \oint\frac{dw}{2\pi iw}\,\log \mathfrak{G}(1/x,w).
\end{equation}
The first integral is a just a constant that will cancel between the $x^+$ and $x^-$ contributions in (\ref{rest-of-crossing}), and only the last integral survives:
\begin{equation}\label{last-int-GG}
 \Delta \tilde{\chi}\equiv  \tilde{\chi}(x^+)+\tilde{\chi}(1/x^+)-\tilde{\chi}(x^-)-\tilde{\chi}(1/x^-)= \frac{2}{i} \oint\frac{dw}{2\pi iw}\,\log \frac{\mathfrak{G}(1/x^-,w)}{\mathfrak{G}(1/x^+,w)}\,.
\end{equation}

\begin{figure}[t]
 \centerline{\includegraphics[width=15cm]{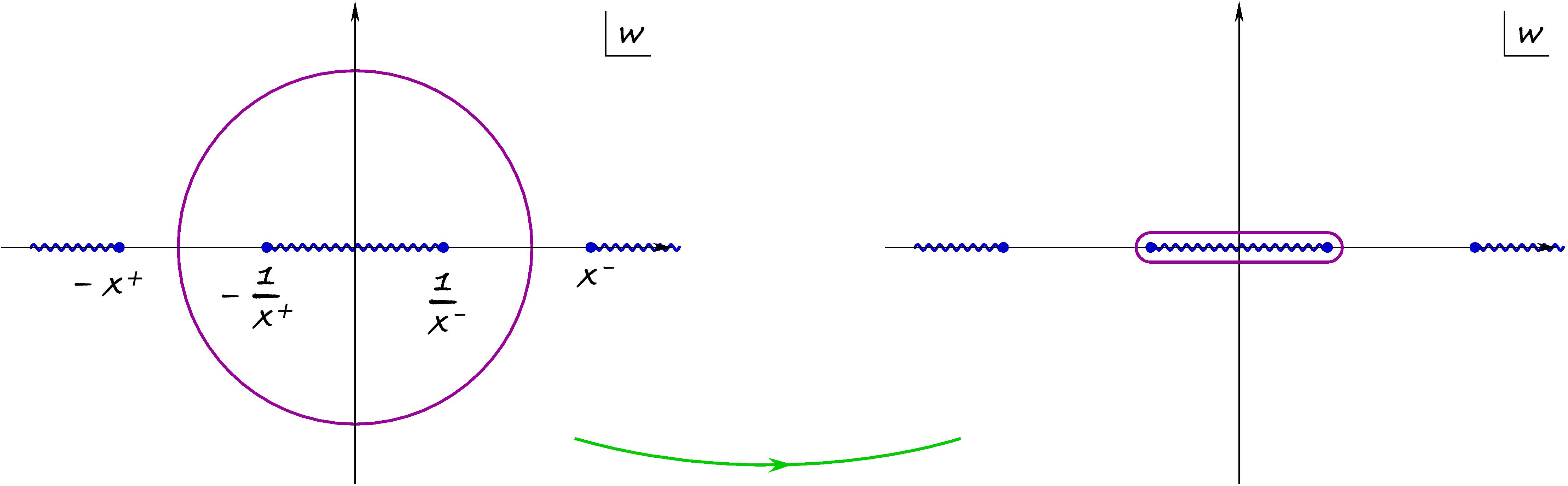}}
\caption{\label{log-cut-picture}\small The contour of integration  in (\ref{last-int-GG}) and the cuts of the integrand.}
\end{figure}

Recalling the definition (\ref{Gamma-functions}) we find for the ratio of the $\mathfrak{G}$-functions:
\begin{align}
 \frac{\mathfrak{G}(1/x^-,w)}{\mathfrak{G}(1/x^+,w)}&=
 \frac{\Gamma \left(\frac{3}{2}+ig\left(x+\frac{1}{x}+w+\frac{1}{w}\right)\right)}{\Gamma \left(\frac{1}{2}-ig\left(x+\frac{1}{x}-w-\frac{1}{w}\right)\right)}
 \,\,
  \frac{\Gamma \left(\frac{3}{2}-ig\left(x+\frac{1}{x}-w-\frac{1}{w}\right)\right)}{\Gamma \left(\frac{1}{2}+ig\left(x+\frac{1}{x}+w+\frac{1}{w}\right)\right)}
\nonumber \\
 &=(ig)^2\left(w+\frac{1}{w}+x^-+\frac{1}{x^-}\right)\left(w+\frac{1}{w}-x^+-\frac{1}{x^+}\right)
\end{align}
The integrand in  (\ref{last-int-GG}) has two log-cuts that have to be chosen as in fig.~\ref{log-cut-picture}, to make the function single-valued on the  contour of integration. Collapsing the contour as shown in the figure picks up the discontinuity of the logarithm equal to $2\pi i$, so
\begin{equation}
  \Delta \tilde{\chi}=\frac{2}{i}\left(
  2\pi i \int_{-1/x^+}^{1/x^-}\frac{dw}{2\pi iw}+2\ln(ig)
  \right)=\frac{2}{i}\ln\left(g^2\,\frac{x^+}{x^-}\right)
\end{equation}
and
\begin{equation}
 \frac{1}{g^4}\,\,{\rm e}\,^{i  \Delta \tilde{\chi}}=\frac{1}{g^4}\,\times \left(g^2\,\frac{x^+}{x^-}\right)^2
\end{equation}
giving precisely (\ref{rest-of-crossing}) as advertised.

\subsection{Asymptotic one-point function}\label{ap:asym}
The asymptotic one-point function for the defect one-point function worked out in \cite{Gombor:2020auk,Gombor:2020kgu} is given by a sum over the integer $is(=g (x_s+1/x_s))$, but here we only need the result for a fixed $s$ and the limit $x_s\to \infty$. The result for a fixed $s$ reads\footnote{Here we omitted an universal overall factor $x_s^{J}$ and $1/\sqrt{J}$.}
\beq
\begin{aligned}
&\frac{\langle D_{s}|\mathcal{O}\rangle}{\sqrt{\langle \mathcal{O}|\mathcal{O}\rangle}}=\\
&\sqrt{\prod_{j=1}^{K_4}f_{\rm SU(2)}(u_{4j})\prod_{j=1}^{K_3}\frac{(x_{3j}^{2}-x_s^{2})}{x_{s}x_{3j}}\prod_{j=1}^{K_5}\frac{(x_{5j}^{2}-x_s^{2})}{x_{s}x_{5j}}\prod_{j=1}^{K_2}\frac{g^2}{u_{2j}(u_{2j}+\frac{i}{2})}\prod_{j=1}^{K_6}\frac{g^2}{u_{6j}(u_{6j}+\frac{i}{2})}{\rm Sdet}G}
\end{aligned}
\eeq
The prefactor $f_{\rm SU(2)}(u)$ is the reflection phase for the operators in the SU(2) sector, which reads \cite{Komatsu:2020sup}
\beq
f_{\rm SU(2)}(u)=\frac{u(u-\frac{i}{2})}{(u-i(s-\frac{1}{2}))(u+i(s-\frac{1}{2}))}\frac{x^{+}}{x^{-}}\frac{x_s^{2}\sigma_B^{D}(u)}{\sigma(u,\bar{u})}\period
\eeq

To obtain the result for the one-point function in the Coulomb branch, we perform the following substitutions to the result above:
\begin{enumerate}
\item Introduce the roots on the first and the last nodes by substituting some of $x_{3j}$ and $x_{5j}$ with $1/x_{1j}$ and $1/x_{7j}$.
\item Set $K_4=K_{1}+K_{3}=K_{5}+K_{7}=2K_{2}=2K_{6}$ in order to satisfy the symmetry constraint. (This constraint comes from the fact that the operator needs to be $SO(5)$ singlet in order to have a non-zero one-point function.)
\item Replace $\sigma_B^{D}$ with $\sigma_B^{D} \sigma_{\rm CDD}$ as discussed above.
\item Take the limit $x_s\to \infty$.
\end{enumerate}
After taking the limit and simplifying the resulting expression using the fact that all the rapidities come in pairs $(u, \bar{u})$, we obtain \eqref{exact-1pt0}.

\section{Descendants in Heisenberg model}\label{OD-Heisenberg}

To elucidate the role of descendants we study in this appendix the boundary states in the $SU(2)$ Heisenberg spin chain. The descendant overlaps in the Heisenberg model are known from first principles, together with their symmetry factors, and we can study how they depend on the quantum nymbers in all the detail. 

The most general eigenstate of the Heisenberg Hamiltonian is an $n$-fold descendant of a Bethe eigenstate:
\begin{equation}\label{descendantsS-}
 \left|\mathbf{u},n\right\rangle=\left(S^-\right)^n\left|\mathbf{u}\right\rangle,
\end{equation}
where $S^-$ is the spin lowering operator and $n$ varies between zero and $2s$ where $s$ is the spin of the multiplet.

We consider two types of integrable boundary states, the matrix product state (MPS)  \cite{deLeeuw:2015hxa}:
\begin{equation}
 \left\langle {\rm MPS}\right|=\sum_{\left\{s_l\right\}}^{}\mathop{\mathrm{tr}}\sigma ^{s_1}\ldots \sigma ^{s_L}\left\langle s_1\ldots s_L\right|,
\end{equation}
where $s_l=\uparrow,\downarrow$ and $\sigma^\uparrow=\sigma _1 $, $\sigma ^\downarrow=\sigma _2$, and the generalized dimer\footnote{VBS stands for the "Valence Bond State".}:
\begin{equation}
\left\langle {\rm VBS}_{\kappa\beta } \right|=\left\langle D_{\kappa\beta } \right|^{\otimes \frac{L}{2}},\qquad 
 \left\langle D_{\kappa\beta } \right|=(\cos\beta +\kappa )\left\langle \uparrow\downarrow\right|+(\cos\beta -\kappa )\left\langle \downarrow\uparrow\right|+i\sin\beta \left(\left\langle \uparrow
 \uparrow\right|+\left\langle \downarrow\downarrow\right|\right).
\end{equation}

These states  are integrable and their overlaps with  Bethe eigenstates are non-vanishing only if the Bethe roots are paired, whereupon the overlaps admit the following determinant representations:
\begin{equation}\label{MPS}
 \frac{\left\langle {\rm MPS}\right.\!\left|\mathbf{u},n \right\rangle}
 {\left\langle \mathbf{u},n\right.\!\left|\mathbf{u},n \right\rangle^{\frac{1}{2}}}
 =\mathbbm{C}_{nK}^{\rm MPS}\cdot 2\,\sqrt{\frac{Q\left(\frac{i}{2}\right)}{Q(0)}\,\,
 \frac{\det G^+}{\det G^-}}\,,
\end{equation}
for the MPS and \cite{Pozsgay:2018ixm,Gombor:2021uxz}
\begin{equation}\label{VBS}
 \frac{\left\langle {\rm VBS}_{\kappa \beta }\right.\!\left|\mathbf{u},n \right\rangle}
 {\left\langle \mathbf{u},n\right.\!\left|\mathbf{u},n \right\rangle^{\frac{1}{2}}}
 =\mathbbm{C}_{nK}^{{\rm VBS}}\cdot \left(\sin\beta \right)^{\frac{L-2K}{2}}\sqrt{\frac{Q^2\left(\frac{i\kappa }{2}\right)}{Q(0)Q\left(\frac{i}{2}\right)}\,\,
 \frac{\det G^+}{\det G^-}}\,,
\end{equation}
for the dimer. The Gaudin matrices are defined by  (\ref{1-loop-Gaudin}) with  $M=2$ for the Catran matrix and  $q=1$ for the Dynkin label. 

The symmetry factors in these two cases are known explicitly \cite{deLeeuw:2017dkd}\footnote{The MPS is only defined for spin chains of an even length, and for an even number of down spins, so the arguments of  the factorials are all integer.}:
\begin{equation}\label{CMPS}
 \mathbbm{C}_{nK}^{\rm MPS}=\frac{\left(\frac{L-2K}{2}\right)!}{\left(\frac{n}{2}\right)!\left(\frac{L-2K -n}{2}\right)!}\sqrt{\frac{n!(L-2K -n)!}{(L-2K)!}}\,,
\end{equation}
and  \cite{Gombor:2021uxz,Kristjansen:2021xno}
\begin{equation}\label{CVBS}
 \mathbbm{C}_{nK}^{\rm VBS}=
 C^{-\frac{L-2K }{2}}_n(i\cot\beta )\sqrt{\frac{n!(L-2K -n)!}{(L-2K) !}}\,,
\end{equation}
where $C^\alpha _n(x)$ are the Gegenbauer polynomials. 

\begin{figure}[t]
\begin{center}
 \centerline{\includegraphics[width=8cm]{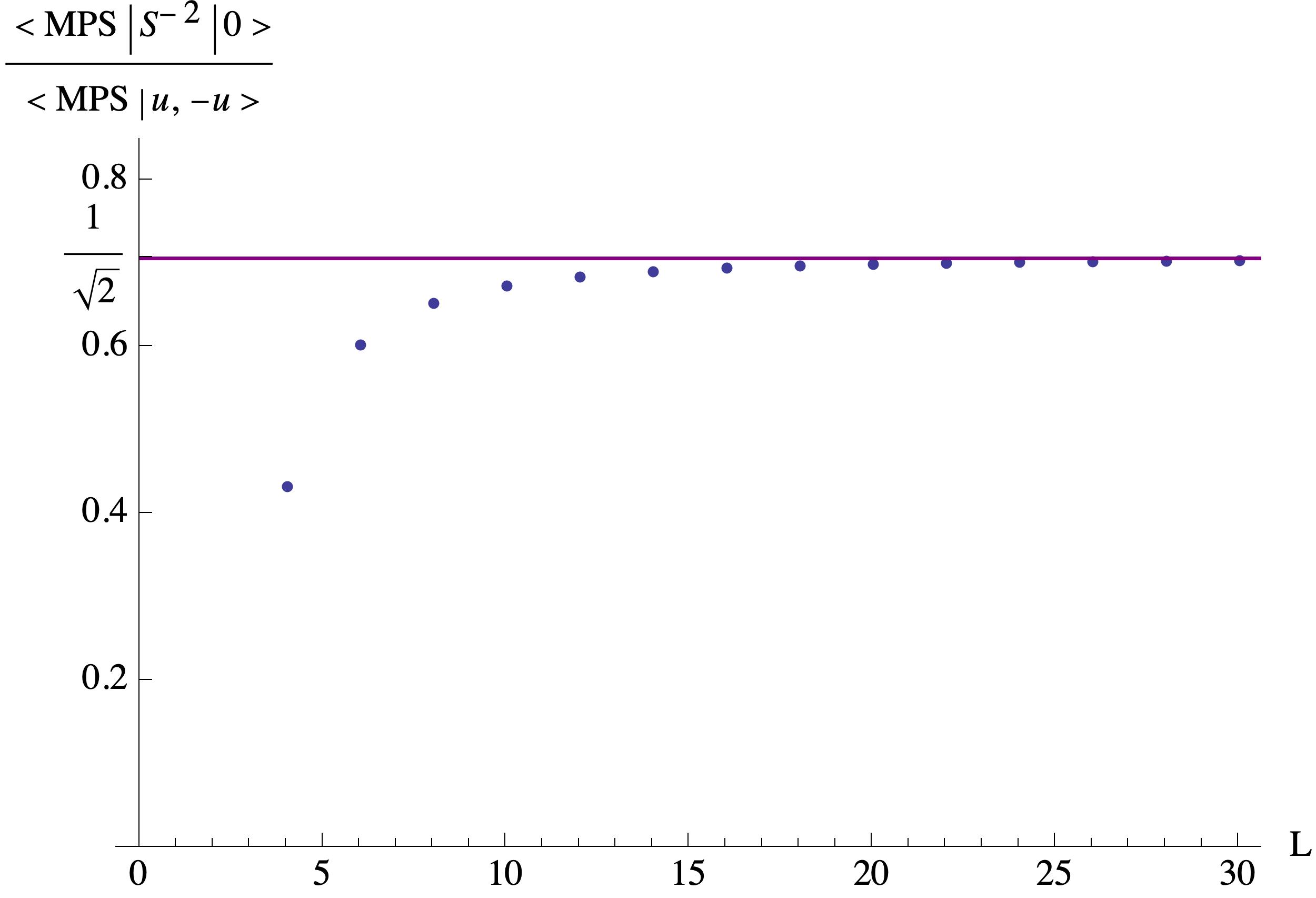}}
\caption{\label{des}\small The ratio of overlaps of the vacuum descendant and the two-magnon state with $u=\frac{1}{2}\,\cot \frac{\pi }{L-1}$, the largest rapidity consistent with momentum quantization. The overlaps are computed directly by constructing the spin states on a chain of finite length.
At large enough length, when $u$ tends to infinity, the ratio approaches $1/\sqrt{2}$ as prescribed by the overlap formula.}
\end{center}
\end{figure}

We start with comparing the overlaps with static magnons (the true descendants) and with very slow magnons. 
While slow magnons approximate descendants arbitrarily well locally, globally a subtle difference always remains. And because the boundary states are long-range entangled the limit of zero momentum is discontinuous. This can be immediately seen from the Bethe-Ansatz formulas above. Eliminating the row and column of the infinite root from the Gaudin matrices (\ref{1-loop-Gaudin}), we find:
\begin{equation}\label{limdet}
 \lim_{v\rightarrow \infty }\frac{\det G^+\left(\left\{u_j,v\right\}\right)}{\det G^-\left(\left\{u_j,v\right\}\right)}
 =\frac{L-2K}{L-2K-1}\,\,\frac{\det G^+\left(\left\{u_j\right\}\right)}{\det G^-\left(\left\{u_j\right\}\right)}\,.
\end{equation}
The overlap of the true descendant, however, is given by (\ref{CMPS}), for MPS. The combinatorial coefficient is then smaller by a factor of $\sqrt{2}$:
\begin{equation}
 \mathbbm{C}_{2K}^{\rm MPS}=\sqrt{\frac{1}{2}\,\,\frac{L-2K}{L-2K-1}}\,.
\end{equation}
This is not a bug.  The ratio of the vacuum descendant and two-magnon overlaps (computed without any recourse to Bethe Ansatz) is plotted in fig.~\ref{des} and shows clear saturation at $1/\sqrt{2}$ for large rapidity (equivalent to small momentum). 

As demonstrated by (\ref{CVBS}), the combinatorial prefactor can be a fairly complicated function of the quantum numbers,  and obviously cannot be obtained by a limiting procedure from the Gaudin factor alone. However, a common feature of the symmetry coefficients (\ref{CMPS}) and (\ref{CVBS}), which they actually share with the limiting form of the Gaudin ratio (\ref{limdet}), is that $K$ only enters in the combination
\begin{equation}
 \nu =L-2K.
\end{equation}
Even if the limiting form of the Gaudin superdeterminant in (\ref{limdet}) does not recover the correct symmetry factor, it is instructive to see how the dependence on $\nu $ arises there.

The dependence on $\nu $ comes from the  diagonal element of the Gaudin matrix
$\partial \chi _j/\partial u_j $, where $\chi _j$ is the phase of the Bethe equation, see (\ref{1loop-BAE}). When $u_j$ becomes very large the phase behaves as $1/u_j$,
and
\begin{equation}\label{nuHeisenberg}
 \nu \equiv \lim_{v\rightarrow \infty }v^2\,\frac{\partial \chi (v)}{\partial v}\,.
\end{equation}
For a state with $K$ excitations $\nu =L-2K$, and this is how $L-2K$ arises in (\ref{limdet}). 

Physically, $\nu $ can be identified with the number  the number of holes  \cite{Faddeev:1996iy} for virtual Bethe roots. The Bethe equation for a very large root is approximated by $\,{\rm e}\,^{i\nu /u}=1$, leaving $\nu $ possible vacancies. And indeed, $L/2-K$ is the spin of the Bethe state, meaning that it belongs to a  representation of dimension $L-2K+1$ so that $n$ in (\ref{descendantsS-}) can take values up  to $L-2k=\nu $.

In a more general situation of the nested Bethe Ansatz of AdS/CFT, we use (\ref{nuHeisenberg}) to define the number of vacancies, that is, the maximal number of virtual roots the state can accommodate. By analogy with the $SU(2)$ overlaps, we conjecture that overlaps of descendants depend exactly on this parameter, and on the number $n$ of virtual roots.

\bibliographystyle{nb}
\bibliography{CoulombRef}	
\end{document}